\documentclass[]{tMPH2e}
\usepackage{xcolor}
\usepackage{tikz}
\definecolor{dgreen}{rgb}{0,.5,0}
\definecolor{dred}{rgb}{.7,.0,.0}
\def\etal{{\it et al.}}

\def\ddroit{{\rm d}}

\begin{document}
\doi{10.1080/0026897YYxxxxxxxx}
 \issn{}
\issnp{}
\jvol{00}
\jnum{00} \jyear{2014} 

\markboth{E.~Fromager}{Molecular Physics}

\articletype{Manuscript}

\title{{\itshape 
On the exact formulation of multi-configuration density-functional
theory: electron density versus orbitals occupation 
}
}

\author{
Emmanuel Fromager$^{\ast}$\thanks{$^\ast$Corresponding author.
Email: fromagere@unistra.fr 
\vspace{6pt}}
\\\vspace{6pt}  
{\em{
Laboratoire de Chimie Quantique,
Institut de Chimie, CNRS / Universit\'{e} de Strasbourg,
4 rue Blaise Pascal, 67000 Strasbourg, France
}};\\\vspace{6pt}  
}

\maketitle

\begin{abstract}

The exact formulation of multi-configuration density-functional theory
(DFT) is discussed in this work. As an alternative to range-separated
methods, where electron correlation effects are split in the coordinate
space, the combination of Configuration Interaction methods with orbital
occupation functionals is explored at the formal level through the
separation of correlation effects in the orbital space. When applied to
model Hamiltonians, this approach leads to an exact Site-Occupation
Embedding Theory (SOET). An adiabatic connection expression is
derived for the complementary bath functional and a comparison with
Density Matrix Embedding Theory (DMET) is made. Illustrative results are
given for the simple two-site Hubbard model. SOET is then applied
to a quantum chemical Hamiltonian, thus leading to an exact Complete
Active Space Site-Occupation Functional Theory (CASSOFT) where active electrons are correlated
explicitly within the CAS and the remaining contributions to the
correlation energy are described with an orbital occupation functional.
The computational implementation of SOET and CASSOFT as well as the
development of approximate functionals are left for future
work.

\bigskip

\bigskip

\end{abstract}

\section{Introduction}\label{sec:intro}

The description of strong electron correlation effects 
is still nowadays a challenging problem for both
density-functional theory (DFT) and wavefunction theory (WFT)
communities. Even though Kohn--Sham DFT (KS-DFT) is in
principle exact, standard approximate exchange--correlation 
functionals usually do not enable an adequate description of
multi-configurational systems. On the other hand, the "Gold standard" single-reference 
{\it Coupled Cluster} (CC) method should also be able to model
near-degeneracies but high-order excitations would then be necessary.   
In practice multi-reference (MR) perturbation theories such as the {\it
second order Complete Active Space Perturbation Theory}
(CASPT2)~\cite{caspt2,WCMS:WCMS97} or the
{\it N-Electron Valence state Perturbation Theory}
(NEVPT2)~\cite{nevpt,nevpt2-spinless} are
usually employed. Despite their success, these methods may suffer from
the perturbative description of
the short-range dynamical correlation which is sometimes not accurate enough.  
For these reasons multi-configurational extensions of both DFT and CC
have been investigated for many years and despite significant efforts
and encouraging results,
it is still unclear how the incorporation of a {\it Complete Active Space Self-Consistent Field}
(CASSCF) into DFT and CC should be performed. A recent review article by Bartlett and
coworkers~\cite{doi:10.1021/cr2001417} gives an extensive picture of the developments in MRCC. We
shall focus here on DFT and restrict ourselves to the description of the
ground state.\\    

The rigorous combination of CASSCF with DFT is already difficult at the
formal level due to the infamous double counting problem. Specific
functionals must indeed be developed for complementing the CASSCF energy
that already contains some correlation
effects~\cite{DoubleCountothers1,DoubleCountothers2,PhysRevA.75.012503,JCP08_Gorling_mcoep,doi:10.1080/00268970903160597}. The latter are usually
referred to as static correlation effects. Only the complementary correlation,
known as dynamical correlation, should therefore be assigned to the
density functional. Turning such a scheme into a practical computational
method is not an easy task since approximations used for standard
functionals cannot be applied straightfowardly in this context.
One way to overcome the double counting problem consists in separating
correlation effects in the coordinate space, as initially proposed by
Savin~\cite{savinbook}. The resulting range-separated DFT methods combine
rigorously long-range WFT with short-range DFT. Even though range
separation allows for a multi-configurational description of the
electron density, it cannot completely isolate static correlation from dynamical
correlation simply because the former is usually not a purely long-range
effect, even in a dissociated molecule. This point will be discussed
further in the following.\\

Let us mention that Savin earlier combined Configuration Interaction
(CI) with DFT
by separating correlation effects in the natural orbital
space~\cite{ijqc88_Andreas_CIDFT_natorb}. Orbitals
with occupation numbers larger than a given
threshold $\nu$ were correlated explicitly, at the CI level, while the
remaining correlation energy was described with a complementary $\nu$-dependent
density functional. More recently, Gutl\'{e} and
Savin~\cite{PhysRevA.75.032519} proposed alternative CI-DFT schemes
where the correlation energy is still split in the orbital space but
through the introduction of gap shift or cutoff parameters. In all these
hybrid CI-DFT schemes, a CI energy is complemented by a functional of
the electron density. In this work we propose to revisit at the formal level the separation of correlation
effects in the orbital space. As a major difference with the
approaches
discussed previously, orbitals occupation will
be used as basic variable rather than the electron density. The idea
originates from condensed matter physics where strongly correlated
electrons are usually described with model
Hamiltonians (such as the Hubbard Hamiltonian) rather than the true physical 
one~\cite{Capelle201391}. As discussed further in the following, adapting such an
approach to quantum chemistry is appealing since static correlation is
usually defined in the orbital space. Molecular orbitals can then be viewed as
sites with an occupation to be determined. As shown in this work, a WFT
description of static correlation can be combined rigorously with 
an orbital-occupation functional modeling of dynamical correlation.
A potential drawback of such a scheme lies in the fact that, in contrast to range-separated DFT,
the complementary functional loses its universality since it depends on
the molecular orbitals. The theory is derived for a particular choice of orbitals,
namely those obtained by diagonalizing the non-interacting Hamiltonian
(kinetic and nuclear potential energy operators only are considered).
Its generalization is left for future work. 
\\   

The paper is organized as follows: After a short introduction to the
double counting problem in multi-configuration DFT
(Sec.~\ref{sec:2blecounting}), the exact multi-determinant extension of
standard DFT based on range separation is briefly reviewed (Sec.~\ref{subsec:srdft}) and its application to the dissociated H$_2$
molecule is discussed in
Sec.~\ref{subsec:H2_min_basis}. We then leave the framework of DFT by
using orbitals occupation as basic
variable rather than the electron density. This approach is first introduced for model Hamiltonians such as the Hubbard
Hamiltonian (Sec.~\ref{sec:SOFT}). By isolating an impurity site and
separating its on-site repulsion from the interactions on the
remaining sites (referred to as the bath), 
an exact {\it Site-Occupation Embedding Theory} (SOET) is derived
and compared with the {\it Density Matrix Embedding Theory} (DMET) of
Knizia and Chan~\cite{PhysRevLett.109.186404}. Illustrative results are then given for the simple two-site
Hubbard model in Sec.~\ref{sec:two_site_Hubb}. Finally, we show in
Sec.~\ref{sec:SOFT_in_QC} that an exact combination of CI with orbital
occupation functional theory is obtained 
when
applying SOET to a quantum chemical Hamiltonian. Conclusions and
perspectives are given
in Sec.~\ref{sec:conclusions}.     
 
\section{The double counting problem}\label{sec:2blecounting}

In WFT the exact ground-state energy of an electronic system can be
obtained variationally as follows,
\begin{eqnarray}\label{eq:exactVP_wft}
E=\underset{\Psi}{\rm
min}\,\langle\Psi\vert\hat{T}+\hat{V}_{\rm ne}+{\hat{W}_{\rm
ee}}\vert\Psi\rangle,
\end{eqnarray}
where $\hat{T}$ is the kinetic energy operator, $\hat{W}_{\rm ee}$
denotes the two-electron repulsion operator and $\hat{V}_{\rm ne}=
\int \ddroit{\bf r}\,v_{\rm ne}
({\bf r})\,\hat{n}({\bf r})
$ is the nuclear potential operator. In a regular CASSCF
calculation,
the minimization in Eq.~(\ref{eq:exactVP_wft}) is restricted to linear combinations of Slater
determinants that belong to a given {\it Complete Active Space} (CAS). The orbitals are also optimized
variationally. The CAS is obtained by distributing a given number of active electrons in
selected active orbitals that become consequently partially occupied. The selection of active electrons and orbitals
is usually based on chemical intuition. Doubly
occupied orbitals are referred to as inactive orbitals. Note that the
latter are not frozen in a CASSCF calculation. The remaining orbitals,
that are not occupied, are the virtuals. 
For convenience we will denote $\mathcal{S}^{M}$ the
space of trial CASSCF wavefunctions. The superscript $M$ specifies
all the restrictions in the minimization (number of active
electrons and active orbitals for example). The CASSCF energy can therefore be written as  
\begin{eqnarray}\label{casscf_ener_exp}
E_{\rm CASSCF}=\underset{\Psi\in \mathcal{S}^{M}}{\rm
min}\,\langle\Psi\vert\hat{T}+\hat{V}_{\rm ne}+{\hat{W}_{\rm
ee}}\vert\Psi\rangle.
\end{eqnarray}
Note that, in practice, a trial CASSCF wavefunction $\Psi$ 
will be parametrized in second quantization as
follows~\cite{casscf_pinkbook}, 
\begin{eqnarray}
\forall \Psi\in\mathcal{S}^{M},\hspace{0.2cm}
\vert \Psi\rangle=e^{-\hat{\kappa}}\Bigg(\sum_{I\in{\rm CAS}} C_I \vert {\rm
det}_I\rangle\Bigg),
\end{eqnarray}
where $\{{\rm det}_I\}_{I\in{\rm CAS}}$ is the basis of determinants for the CAS. The singlet excitation operator $\hat{\kappa}$ that
allows for orbital rotations is defined
as
\begin{equation}
\hat{\kappa}=\sum_{p>q,\sigma=\uparrow,\downarrow}\kappa_{pq}\Big(\hat{a}^{\dagger}_{p\sigma}\hat{a}_{q\sigma}-\hat{a}^{\dagger}_{q\sigma}\hat{a}_{p\sigma}\Big),
\end{equation}
where $p$ and $q$ can be inactive, active or virtual
orbitals. Consequently, the minimization in
Eq.~(\ref{casscf_ener_exp}) is performed over both
$\{\kappa_{pq}\}_{p>q}$ and $\{C_I\}_{I\in{\rm CAS}}$ parameters.\\

The energy difference $E-E_{\rm CASSCF}$ is usually referred to as
dynamical correlation energy. The latter is routinely described within multi-reference
perturbation theories such as CASPT2 or
NEVPT2. 
A combined CASSCF-DFT approach should ideally be exact and variational.
Dynamical correlation would be described with a density functional
$\overline{E}^{{{\rm dyn}}}_{\rm c}[n]$ that would complement the CASSCF
energy. This can be formulated rigorously when considering the universal
Levy--Lieb functional~\cite{LFTransform-Lieb}, 
\begin{eqnarray}\label{eq:universal_LL_fun}
F[n] = 
\underset{\Psi\rightarrow n}{\rm min}\langle \Psi\vert
\hat{T}+{\hat{W}_{\rm ee}}\vert\Psi\rangle,
\end{eqnarray}
where the minimization is performed over all wavefunctions with density
$n$, and the following partitioning~\cite{DoubleCountothers2},
\begin{eqnarray}\label{LL_CAS_fun}
F[n] =F^{M}[n]+\overline{E}^{{{\rm dyn}}}_{\rm c}[n]. 
\end{eqnarray}
The LL functional associated with the CASSCF space $\mathcal{S}^{M}$ is defined
as
\begin{eqnarray}
F^{{{M}}}[n] = 
\underset{\Psi\in\mathcal{S}^{M}\rightarrow n}{\rm min}\langle \Psi\vert
\hat{T}+{\hat{W}_{\rm ee}}\vert\Psi\rangle,
\end{eqnarray}
where the minimization is restricted to CASSCF wavefunctions in
$\mathcal{S}^{M}$ with density $n$.
According to the Hohenberg--Kohn (HK) theorem~\cite{hktheo}, the exact ground-state energy can be
obtained variationally as follows,
\begin{eqnarray}\label{eq:HK2_theo}
E&=&\underset{n}{\rm min}\Big\{
F[n]+
(v_{\rm ne}\vert n)
\Big\},
\end{eqnarray}
where the minimization is performed over electron densities $n({\bf r})$ that
integrate to a fixed number $N$ of electrons. The notation $(v\vert n)=
\int \ddroit{\bf r}\,v
({\bf r})\,n({\bf r})
$ has been used. Combining Eq.~(\ref{LL_CAS_fun}) with
Eq.~(\ref{eq:HK2_theo}) leads to 
\begin{eqnarray}\label{eq:casdft_min_n}
E&=&\underset{n}{\rm min}\Big\{
F^{{{M}}}[n]+\overline{E}^{{{\rm dyn}}}_{\rm c}[n]+
(v_{\rm ne}\vert n)
\Big\}.
\end{eqnarray}
Consequently, for any trial CASSCF wavefunction $\Psi$ in
$\mathcal{S}^{M}$,
\begin{eqnarray}
\langle\Psi\vert\hat{T}+\hat{V}_{\rm ne}+{\hat{W}_{\rm
ee}}\vert\Psi\rangle+{\overline{E}^{{{\rm dyn}}}_{\rm
c}[n_{\Psi}{]}}&\geq& F^{{{M}}}[n_{\Psi}]+{\overline{E}^{{{\rm
dyn}}}_{\rm c}[n_{\Psi}{]}}+(v_{\rm ne}\vert n_\Psi)
\nonumber\\
&\geq& E,
\end{eqnarray} 
where $n_{\Psi}({\bf r})=\langle \Psi \vert \hat{n}({\bf r})\vert
\Psi\rangle$ is the electron density obtained from the trial CASSCF
wavefunction $\Psi$. 
Thus we conclude that, provided that the exact ground-state density can be
reproduced by a CASSCF wavefunction, the exact ground-state energy can
be written as  
\begin{eqnarray}\label{eq:VAR_casdft_ener}
E=\underset{\Psi\in \mathcal{S}^{M}}{\rm
min}\,\Big\{
\langle\Psi\vert\hat{T}+\hat{V}_{\rm ne}+{\hat{W}_{\rm
ee}}\vert\Psi\rangle+{\overline{E}^{{{\rm dyn}}}_{\rm
c}[n_{\Psi}{]}}
\Big\}.
\end{eqnarray}
In order to turn Eq.~(\ref{eq:VAR_casdft_ener}) into a practical
computational method, approximate complementary correlation functionals
should be developed. Obviously standard correlation functionals cannot
be used otherwise correlation effects would be double counted. This
double counting problem arises also when the size of the CAS varies. In
the exact theory, the functional should vary with the CAS so that
correlation effects are transferred from the CASSCF to the DFT part of
the energy and the total energy remains constant and equal to the exact
ground-state energy. Deriving CAS-dependent density functionals is a
complicated task simply because the electron density is defined in the
coordinate space while a CAS is defined in the orbital space. In order
to rigorously overcome such difficulties, Savin~\cite{savinbook} proposed to separate
correlation effects in the coordinate space, thus leading to the
so-called range-separated DFT methods. This approach is briefly reviewed
in Sec.~\ref{sec:MC-srDFT}. We then propose an alternative
approach where correlation effects are separated in the orbital space,
like in a regular CASSCF calculation. In this context, the orbitals occupation will be
used as basic variable rather than the electron density.           

\section{Multi-configuration range-separated DFT}\label{sec:MC-srDFT}

\subsection{Separating correlations in coordinate
space}\label{subsec:srdft}

In the standard KS-DFT scheme~\cite{kstheo}, the universal LL functional
in Eq.~(\ref{eq:universal_LL_fun}) is decomposed into
the non-interacting kinetic energy functional $T_{\rm s}[n]=
\underset{\Psi\rightarrow n}{\rm min}\langle \Psi\vert
\hat{T}\vert\Psi\rangle
$ and the Hartree-exchange-correlation (Hxc) energy functional,
\begin{eqnarray}
F[n]=T_{\rm s}[n]+E_{\rm Hxc}[n].
\end{eqnarray}
Consequently, the HK variational principle can be reformulated as
follows,
\begin{eqnarray}
E&=&\underset{\Psi}{\rm min}\Big\{
\langle\Psi\vert\hat{T}+\hat{V}_{\rm ne}\vert\Psi\rangle
+E_{\rm Hxc}[n_\Psi]
\Big\},
\end{eqnarray}
where the minimizing KS determinant $\Phi^{\rm KS}$ fulfils the self-consistent equation
\begin{eqnarray}
\Bigg(\hat{T}+\hat{V}_{\rm ne}+\int \ddroit{\bf r}\,\dfrac{\delta E_{\rm
Hxc}}{\delta n({\bf r})}[n_{\Phi^{\rm KS}}]\,\hat{n}({\bf r})\Bigg)
\vert\Phi^{\rm KS}\rangle=\mathcal{E}^{\rm KS}\vert\Phi^{\rm KS}\rangle.
\end{eqnarray}  
Therefore, within KS-DFT, the two-electron interaction is fully
described by a density functional. As originally shown by
Savin~\cite{savinbook}, it is in fact
possible to describe only a part of the two-electron repulsion within
DFT and leave
the remaining part to WFT. This is achieved by separating the interaction in the coordinate space into
two complementary contributions. The range separation based on the error
function,
\begin{eqnarray}
w_{\rm ee}(r_{12})&=&w^{\rm lr,\mu}_{\rm ee}(r_{12})+w^{\rm sr,\mu}_{\rm
ee}(r_{12}),
\nonumber\\
w^{\rm lr,\mu}_{\rm ee}(r_{12})&=&{\rm erf}(\mu r_{12})/r_{12},
\end{eqnarray}
has for example been used extensively in the last decade (see Ref.~\cite{JChemPhys139_2013}
and the references therein), but any separation like the
simpler linear one~\cite{Sharkas_JCP} can be considered. Range
separation, that is controlled by the $\mu$ parameter, is appealing as it enables to
isolate the Coulomb hole and assign it to a density functional while
long-range correlation can be described in WFT. 
This hybrid range-separated WFT-DFT approach can be derived rigorously from the
alternative partitioning of the LL functional,
\begin{eqnarray}\label{eq:LL_range_separation}
F[n]=
F^{\rm lr,\mu}[n]+\overline{E}^{\rm sr,\mu}_{\rm Hxc}[n],
\end{eqnarray}
where the long-range LL functional equals 
\begin{eqnarray}\label{eq:LL_longrange}
F^{\rm lr,\mu}[n]=
\underset{\Psi\rightarrow n}{\rm min}\langle \Psi\vert
\hat{T}+{\hat{W}^{\rm lr,\mu}_{\rm ee}}\vert\Psi\rangle
,
\end{eqnarray}
and $\overline{E}^{\rm sr,\mu}_{\rm Hxc}[n]$ is the complementary
$\mu$-dependent short-range Hxc density functional. Let us stress that, for any $\mu$ value, this functional is universal since it
depends on the electron density only. Combining Eq.~(\ref{eq:LL_range_separation})
with Eq.~(\ref{eq:HK2_theo}) leads to the exact ground-state energy
expression
\begin{eqnarray}
\displaystyle
E=\underset{\Psi}{\rm min}\left\{ 
\langle \Psi\vert
\hat{T}+\hat{W}^{\rm lr,\mu}_{\rm ee}+\hat{V}_{\rm ne}\vert\Psi\rangle
+\overline{E}^{\rm sr,\mu}_{\rm Hxc}[n_{\Psi}]
\right\},
\end{eqnarray}
where the minimizing wavefunction $\Psi^\mu$ fulfils the self-consistent
equation
\begin{eqnarray}\label{timeindepsrdfteq}
&&\hat{H}^\mu\vert {\Psi}^{\mu}\rangle
={\mathcal{E}}^{\mu}\vert {\Psi}^{\mu}\rangle,
\nonumber\\
&&\hat{H}^\mu
=
\hat{T}+\hat{W}^{\rm lr,\mu}_{\rm
ee}+\hat{V}_{\rm ne}
+
\int \ddroit\mathbf{r}
\dfrac{\delta \overline{E}^{{\rm sr,\mu}}_{\rm
Hxc}}{\delta
n(\mathbf{r})}[n_{\Psi^\mu}]
\hat{n}(\mathbf{r}).
\end{eqnarray}
Note that KS-DFT and pure WFT are recovered when $\mu=0$ and
$\mu\rightarrow+\infty$, respectively.\\
Since the long-range interaction is now described explicitly, the auxiliary
wavefunction $\Psi^\mu$, that has exactly the same density as the
physical system, is multi-determinantal. CASSCF can therefore be applied
in this context in conjunction with local or semi-local short-range
functionals~\cite{JCPunivmu,JCPunivmu2}.
Even though encouraging results were obtained with such a
range-separated CASDFT scheme, better functionals are still needed for
the method to be reliable~\cite{JChemPhys139_2013}. As discussed further in
Sec.~\ref{subsec:H2_min_basis}, one major problem is that static correlation is
not a purely long-range correlation effect.\\

Returning to the exact theory, an adiabatic connection
(AC)~\cite{LANGRETH:1975p1425,GUNNARSSON:1976p1781,Gunnarsson:1977,LANGRETH:1977p1780,SavRev,Yang:1998p441} expression for the short-range
functional can be obtained from the auxiliary
equations,
\begin{eqnarray}
\Bigg(\hat{T}+\hat{W}^{\rm lr,\nu}_{\rm
ee}+
\int \ddroit\mathbf{r}\,
v^{\nu}(\mathbf{r})\hat{n}(\mathbf{r})
\Bigg)\vert {\Psi}^{\nu}\rangle={\mathcal{E}}^{\nu}\vert
{\Psi}^{\nu}\rangle,
\end{eqnarray}
and the density constraint $n_{{\Psi}^{\nu}}(\mathbf{r})=n(\mathbf{r})$
for $0\leq \nu<+\infty$. Indeed, according to
Eqs.~(\ref{eq:LL_range_separation}) and (\ref{eq:LL_longrange}),           
\begin{eqnarray}
\overline{E}^{\rm sr,\mu}_{\rm Hxc}[n]&=&F[n]-F^{\rm lr,\mu}[n]
\nonumber\\
&=&
\int^{+\infty}_\mu \dfrac{\ddroit\mathcal{E}^\nu}{\ddroit \nu}\ddroit\nu\
+(v^\mu-v^{+\infty}\vert n),
\end{eqnarray}
thus leading, according to the Hellmann--Feynman theorem, to
\begin{eqnarray}\label{eq:AC_srfun}
\overline{E}^{\rm sr,\mu}_{\rm Hxc}[n]
&=&\int^{+\infty}_\mu \langle {\Psi}^{\nu}\vert\partial \hat{W}^{\rm lr,\nu}_{\rm
ee}/\partial \nu \vert{\Psi}^{\nu} \rangle.  
\end{eqnarray}
We obtain for $\mu=0$ a range-separated AC
expression for the conventional Hxc functional,
\begin{eqnarray}\label{eq:AC_RS_convHxc}
{E}_{\rm Hxc}[n]
&=&
\int^{+\infty}_0 \langle {\Psi}^{\nu}\vert\partial \hat{W}^{\rm lr,\nu}_{\rm
ee}/\partial \nu \vert{\Psi}^{\nu} \rangle
.  
\end{eqnarray}
It is readily seen from Eqs.~(\ref{eq:AC_srfun}) and
(\ref{eq:AC_RS_convHxc}) that conventional functionals cannot be used
straightfowardly in multi-determinant range-separated DFT otherwise there would be double counting of long-range correlation
effects through the WFT treatment. By connecting the non-interacting KS
system ($\nu=0$) to the
long-range interacting one ($\nu=\mu$) as follows,
\begin{eqnarray}
\int^{\mu}_0 \langle {\Psi}^{\nu}\vert\partial \hat{W}^{\rm lr,\nu}_{\rm
ee}/\partial \nu \vert{\Psi}^{\nu} \rangle&=&
\int^{\mu}_0 \dfrac{\ddroit\mathcal{E}^\nu}{\ddroit \nu}\ddroit\nu\
+(v^0-v^\mu\vert n)
=F^{\rm lr,\mu}[n]-T_{\rm s}[n]
\nonumber\\
&=&{E}^{\rm lr,\mu}_{\rm Hxc}[n],
\end{eqnarray}
where ${E}^{\rm lr,\mu}_{\rm Hxc}[n]$ is the purely long-range Hxc
functional, we obtain from Eqs.~(\ref{eq:AC_srfun}) and
(\ref{eq:AC_RS_convHxc}) the following expression,
\begin{eqnarray}
\overline{E}^{\rm sr,\mu}_{\rm Hxc}[n]={E}_{\rm Hxc}[n]-{E}^{\rm
lr,\mu}_{\rm Hxc}[n],
\end{eqnarray}     
that has been used by Toulouse~\etal~\cite{erferfgaufunc} for developing approximate
local and semi-local short-range functionals. We should stress that all
the formalism briefly reviewed in this section is in fact general and
can be applied in a different context, for example in DFT for model
Hamiltonians, as proposed
in Sec.~\ref{subsec:soet}. 

\subsection{Left-right correlation and range
separation}\label{subsec:H2_min_basis}

We consider in this section the H$_2$ molecule in a Slater minimal basis
consisting of the $1s_A$ and $1s_B$ atomic orbitals localized on the
left and right hydrogen
atoms, respectively~\cite{H2minbasissetJChemEduc,Sharkas_JCP}. The basis functions are identical with $\zeta=1$. For large bond distances the bonding and anti-bonding molecular
orbitals are equal to 
$1\sigma_g=\frac{1}{\sqrt{2}}\big(1s_A+1s_B\big)$
and
$1\sigma_u=\frac{1}{\sqrt{2}}\big(1s_A-1s_B\big)$, respectively. 
The ground state will then be written in the basis the two Slater determinants $1\sigma_g^2$ and
$1\sigma_u^2$. Since the latter differ by a double excitation,
they are not
coupled by one-electron operators. Therefore, when approaching the dissociation limit, the matrix representation of the auxiliary long-range
Hamiltonian in Eq.~(\ref{timeindepsrdfteq}) reduces to
\begin{eqnarray}
[\hat{H}^\mu]=\left[
\begin{array}{c c}
{E}^\mu & K^{{\mu}} \\
{K^{{\mu}}} & {E}^\mu
\end{array}
\right],
\end{eqnarray}
where diagonal elements are identical since atomic orbitals do not
overlap, and the coupling term equals   
\begin{eqnarray}\label{eq:Kmu_exp}
K^{{\mu}} &=& \langle 1\sigma_g^2 \vert {\hat{W}^{\rm
lr,{{\mu}}}_{\rm ee}}\vert  1\sigma_u^2
\rangle=
\langle1\sigma_u1\sigma_u\vert{w^{\rm lr,{\mu}}_{\rm
ee}(r_{12})}\vert 1\sigma_g1\sigma_g\rangle
\nonumber\\
&=&\displaystyle
\frac{1}{2}{
\langle1s_A1s_A\vert{w^{\rm lr,{\mu}}_{\rm
ee}(r_{12})}\vert 1s_A1s_A\rangle}
-\frac{1}{2}{
\langle1s_A1s_B\vert{w^{\rm lr,{\mu}}_{\rm
ee}(r_{12})}\vert 1s_A1s_B\rangle}.
\end{eqnarray}
While the second long-range two-electron integral on the right-hand side of Eq.~(\ref{eq:Kmu_exp})
reduces to the regular one $\langle1s_A1s_B\vert 1s_A1s_B\rangle$ and
becomes zero in the dissociation limit, the first term is an "on-site"
integral computed with the long-range interaction. Thus we conclude that 
the coupling between $1\sigma_g^2$ and $1\sigma_u^2$ determinants
is determined by the contribution at short range of the long-range
interaction. 
In this case static correlation, that is also referred to as left-right
correlation, can obviously not be
interpreted as a purely long-range correlation effect. Nevertheless, if the
error function is used, we see from its Taylor expansion for small ${\mu}r_{12}$,
\begin{eqnarray}
\displaystyle w^{\rm lr,{\mu}}_{\rm ee}(r_{12})
=\frac{2}{\sqrt{\pi}}\left({\mu}-\frac{1}{3}{\mu^3}r_{12}^2+\ldots\right), 
\end{eqnarray} 
that the coupling term can be expanded as follows
\begin{eqnarray}
\displaystyle
K^{{\mu}}=\frac{{\mu}}{\sqrt{\pi}}+\ldots
\end{eqnarray}
Consequently, as already pointed out by Gori-Giorgi and
Savin~\cite{PaolasrXmd},
even an infinitesimal $\mu$ value ensures that $K^{{\mu}}$ is not
strictly equal to zero, thus providing the correct multi-configurational
description of the dissociated H$_2$ molecule in the ground state:
\begin{eqnarray}
\displaystyle\vert\Psi^{{\mu}}\rangle=
\frac{1}{\sqrt{2}}\Big(\vert1\sigma_g^2\rangle-\vert1\sigma_u^2\rangle\Big).
\end{eqnarray}
This simple example illustrates how difficult it is to describe  
static correlation   
in the coordinate space, in contrast to short-range dynamical
correlation that is connected with the Coulomb hole. We propose in the
rest of this paper an alternative approach where correlations are
separated in the orbital space. In order to overcome the
double counting problem, the orbitals occupation will be used
rather than the electron density. For clarity, this approach will be
introduced first for model Hamiltonians.
 
\section{DFT for model Hamiltonians}\label{sec:SOFT}

As mentioned previously, it is convenient to work in the orbital
space rather than the coordinate space when it comes to separate static
and dynamical correlation effects. A change of paradigm is then
necessary in order to avoid the double counting problem. The basic
variable in DFT is the electron density $n(\mathbf{r})$ that is defined
in the coordinate
space. As we want to reformulate DFT in the orbital space, orbitals
occupation $\{n_i\}_i$ seems to be the variable of choice. This is known in condensed matter physics as {\it
Site-Occupation Functional Theory}
(SOFT)~\cite{PhysRevLett.56.1968,PhysRevLett.90.146402,Capelle201391}. The latter is nothing
but the formulation of DFT for model Hamiltonians such as the Hubbard
Hamiltonian. After a short
introduction to the KS-SOFT scheme in Sec.~\ref{subsec:ks-soft}, we will show in
Sec.~\ref{subsec:soet} how formal analogies
with range-separated DFT can lead in this context to an exact
embedding theory 
for model Hamiltonians.    

\subsection{SOFT and its KS formulation}\label{subsec:ks-soft}

Let us consider the Hubbard Hamiltonian with an external
potential ${v}\equiv \{{v_i}\}_i$:  
\begin{eqnarray}\label{eq:Hubb_hamil_ext_pot}
\hat{H}=\hat{\mathcal{T}}+\hat{U}+ 
\sum_iv_i\hat{n}_{i},
\nonumber\\
\hat{\mathcal{T}}=
{-t}\sum_{i\neq
j,\sigma}\hat{a}^\dagger_{i\sigma}\hat{a}_{j\sigma},\nonumber\\
\hat{U}=
{U}\sum_i
\hat{n}_{i\uparrow}\hat{n}_{i\downarrow},
\end{eqnarray}
where $t$ is the hopping integral, $U$ denotes the on-site two-electron
repulsion,
$\hat{n}_{i\sigma}=\hat{a}^\dagger_{i\sigma}\hat{a}_{i\sigma}$ and
$\sigma=\uparrow,\downarrow$. The site-occupation operator equals 
$\hat{n}_{i}=\hat{n}_{i\uparrow}+\hat{n}_{i\downarrow}$ and, for a given
wavefunction $\Psi$, the occupation of site $i$ is defined as $n_i=\langle
\Psi\vert \hat{n}_{i}\vert\Psi\rangle$.
For simplicity we will consider a fixed number of electrons
$N$ and only discuss ground-state properties. 
As shown by Gunnarsson and
Sch\"{o}nhammer~\cite{PhysRevLett.56.1968}, the HK theorem can be adapted to the
Hamiltonian in Eq.~(\ref{eq:Hubb_hamil_ext_pot}). There is indeed a one-to-one correspondence
between the external potential $v$ and the ground-state sites 
occupancy $n\equiv\{n_i\}_i$. Consequently, for fixed $t$ and $U$
parameters, the exact ground-state
energy can be obtained from the following variational principle,   
\begin{eqnarray}\label{eq:VP_soft}
\displaystyle E({v})=\underset{n}{\rm min}\Big\{F(n)+
({v}\vert n)\Big\},
\end{eqnarray}
where 
$
({v}\vert n)=
\sum_i{v_i}\,{n}_i$. The analog of the universal HK functional is a function of
the sites occupation that can be written within the LL
constrained-search formalism as 
\begin{eqnarray}
F(n)=
\underset{\Psi\rightarrow
n}{\rm
min}\left\{\langle\Psi\vert\hat{\mathcal{T}}+\hat{U}\vert\Psi\rangle\right\}.
\end{eqnarray}

%

Since the {HK theorem still holds} when ${U}=0$, a KS formulation of
SOFT (KS-SOFT) is obtained from the following partitioning,
\begin{eqnarray}\label{eq:HxcSOfun_def}
F(n)=\mathcal{T}_{\rm s}(n)+E_{\rm Hxc}(n),
\end{eqnarray} where the non-interacting kinetic energy functional is
defined in analogy with KS-DFT as
\begin{eqnarray}
\mathcal{T}_{\rm s}(n)=
\underset{\Psi\rightarrow
n}{\rm
min}\left\{\langle\Psi\vert\hat{\mathcal{T}}\vert\Psi\rangle\right\}
=\langle\Psi^{\rm KS}(n)\vert\hat{\mathcal{T}}\vert\Psi^{\rm KS}(n)\rangle.
\end{eqnarray}
Note that, in this context, the non-interacting KS system is
{\it not}
described by sites that are strictly singly, doubly or not occupied, 
\begin{eqnarray}
&&\vert\Psi^{\rm
KS}(n)\rangle\neq\prod_{i,\sigma}\Big(\hat{a}^\dagger_{i\sigma}\Big)^{n_{i\sigma}}\vert{\rm
vac}\rangle,\nonumber\\
&&n_{i\sigma}=0,1,\nonumber\\
&&\displaystyle\sum_{i,\sigma}n_{i\sigma}=N.
\end{eqnarray}
In other words $\Psi^{\rm KS}(n)$ is not a single Slater determinant in
the basis of the site orbitals. Indeed, $\Psi^{\rm KS}(n)$ is the
ground state of the non-interacting Hamiltonian
$\hat{\mathcal{T}}+\sum_i{v^{\rm
KS}_i}(n)\,\hat{n}_i$ where
$\hat{\mathcal{T}}$ is {\it non-local} in the orbital space.  
Let us stress that there is {no HK theorem without a non-local term}
like $\hat{\mathcal{T}}$ 
included into the Hamiltonian. For example, for two electrons, the two
Hamiltonians that differ by more than a constant,
\begin{eqnarray}
\varepsilon_1\hat{n}_1+\varepsilon_2\hat{n}_2,
\end{eqnarray} 
 and
\begin{eqnarray}
\varepsilon_1\hat{n}_1+\big(\varepsilon_2+\delta\big)\hat{n}_2,
\end{eqnarray} 
where $\varepsilon_1<\varepsilon_2$ and $\delta>
\varepsilon_1-\varepsilon_2$, {yield
the same ground-state occupancies} $n_1=2$
and $n_2=0$.  
This is an important difference with {KS-DFT} where the {quantity to
reproduce} from singly or {doubly occupied} KS orbitals  
is the exact electron {density} $n(\mathbf{r})$.
In {KS-SOFT}, the quantity to reproduce is the exact sites occupation
$n$ that may be fractional due to the non-zero hopping integral
${t}$.\\

By analogy with KS-DFT, an exact AC expression for the Hxc site-occupation
functional in Eq.~(\ref{eq:HxcSOfun_def}) can be obtained from the
following auxiliary
equations,
\begin{eqnarray}\label{eq:AC_full_scale}
\Big(\hat{\mathcal{T}}+\lambda\hat{U}
+\sum_iv^\lambda_i\hat{n}_{i}\Big)\vert\Psi^\lambda\rangle=\mathcal{E}^\lambda\vert\Psi^\lambda\rangle
,
\end{eqnarray}
with the site-occupation constraint $n_{\Psi^\lambda}=n$ fulfilled for
$0\leq \lambda\leq 1$, thus leading to 
\begin{eqnarray}\label{eq:Hxc_fun_ACexp}
E_{\rm Hxc}(n)&=&
\int^1_0 \ddroit \lambda \dfrac{\ddroit
\mathcal{E}^\lambda}{\ddroit \lambda}+(v^0-v^1\vert n)
\nonumber\\
&=&\int^1_0 \ddroit \lambda\;
\langle \Psi^\lambda \vert \hat{U}\vert
\Psi^\lambda\rangle
,
\end{eqnarray}
according to the Hellmann--Feynman theorem. Note that within the mean-field approximation and in the particular case
of a singlet ground state, the Hxc functional is simplified as follows, 
\begin{eqnarray}\label{MF_approx_Hxc}
\displaystyle {E}_{\rm Hxc}(n)\rightarrow
{U}\sum_i{n}_{i\uparrow}{n}_{i\downarrow}=
\dfrac{{U}}{4}\sum_i{n}^2_{i}.
\end{eqnarray}
The latter expression is usually referred to as the Hartree energy in condensed
matter physics~\cite{Capelle201391}. The exchange energy is then considered to be zero.



\subsection{Exact embedding within SOFT}\label{subsec:soet}

In the spirit of
{\it Dynamical Mean-Field
Theory}~\cite{RevModPhys96_dmft_Georges,RevModPhys06_dmft_Kotliar} and
DMET~\cite{PhysRevLett.109.186404}, we propose to isolate one 
site, referred to as {impurity} and for which correlation effects will be 
described explicitly, while the remaining sites (the bath) are treated
within SOFT. Such an approach, that will be referred to as {\it Site-Occupation
Embedding Theory} (SOET) in the following, is formally very similar to range-separated DFT.
The main difference is that the interactions are separated here in the
orbital space rather than in the coordinate space. For convenience 
the impurity and bath sites will be labelled as $i=0$ and $i>0$,
respectively. By analogy with Eqs.~(\ref{eq:LL_range_separation}) and
(\ref{eq:LL_longrange}),
SOET can be derived from the following partitioning of the LL functional,
\begin{eqnarray}\label{eq:Fimp_plus_Ebath}
F(n)=F^{\rm imp}(n)+\overline{E}^{\rm bath}_{\rm Hxc}(n),
\end{eqnarray}  
where the impurity LL functional equals 
\begin{eqnarray}\label{eq:LLimpfun}
F^{\rm imp}(n)=\underset{\Psi\rightarrow
n}{\rm
min}\left\{
\langle\Psi\vert\hat{\mathcal{T}}+{U}\hat{n}_{0\uparrow}\hat{n}_{0\downarrow}\vert\Psi\rangle
\right\},
\end{eqnarray}
and $\overline{E}^{\rm bath}_{\rm Hxc}(n)$ denotes the complementary Hxc
bath functional that describes the bath as well as the coupling between
the bath and the impurity. 
Note that, in contrast to range-separated DFT, the separation of
correlation effects is, in SOET, not
controlled by a single continuous parameter. It rather relies on the
selection of an impurity site. As discussed in the following more
impurity sites can actually be considered.\\
 
We now return to the single-impurity
case. Since, according to Eqs.~(\ref{eq:VP_soft}),
(\ref{eq:Fimp_plus_Ebath}) and (\ref{eq:LLimpfun}), any
normalized trial wavefunction $\Psi$ fulfils
\begin{eqnarray}
&&\langle\Psi\vert\hat{\mathcal{T}}+{U}\hat{n}_{0\uparrow}\hat{n}_{0\downarrow}\vert\Psi\rangle
+\overline{E}^{\rm bath}_{\rm Hxc}(n_\Psi)+({v}\vert n_\Psi)\nonumber\\
&&\geq {F^{\rm
imp}(n_\Psi)+
\overline{E}^{\rm bath}_{\rm Hxc}(n_\Psi)}
+({v}\vert n_\Psi)\geq E({v}),
\end{eqnarray}
the exact ground-state energy can be expressed as 
\begin{eqnarray}
E({v})=
\underset{\Psi
}{\rm
min}\Big\{
\langle\Psi\vert
\hat{\mathcal{T}}+{U}\hat{n}_{0\uparrow}\hat{n}_{0\downarrow}
\vert\Psi\rangle+
\overline{E}^{\rm bath}_{\rm Hxc}(n_\Psi)
+({v}\vert n_\Psi)\Big\},
\end{eqnarray}
where the minimizing wavefunction ${\Psi}^{\rm imp}$ fulfils the
self-consistent equation
\begin{eqnarray}\label{eq:SC_eq_imp}
&&\Bigg(
\hat{\mathcal{T}}+{U}\hat{n}_{0\uparrow}\hat{n}_{0\downarrow}+
\sum_i\Bigg[{v_i}+\dfrac{\partial \overline{E}^{\rm bath}_{\rm
Hxc}(n_{{\Psi^{\rm imp}}
})}
{\partial
n_i}\Bigg]\hat{n}_i\Bigg)\vert{\Psi}^{\rm imp}\rangle
\nonumber
\\
&&={\mathcal{E}}^{\rm imp}
\vert{\Psi}^{\rm imp}\rangle. 
\end{eqnarray}
Thus we obtain an exact embedding scheme where the bath is described
by a site-occupation functional. Its connection with DMET will be
discussed in the following. Let us first focus on the complementary bath functional. From the KS
decomposition 
\begin{eqnarray}\label{eq:LL_imp_KSdecomp}
F^{\rm imp}(n)=\mathcal{T}_{\rm
s}(n)+{E}^{\rm imp}_{\rm Hxc}(n),
\end{eqnarray}
where ${E}^{\rm imp}_{\rm Hxc}(n)$ describes the repulsion on
the impurity that is embedded into a non-interacting bath, we obtain with Eqs.~(\ref{eq:HxcSOfun_def}) and
(\ref{eq:Fimp_plus_Ebath}),
\begin{eqnarray}\label{eq:bathfun_as_diff}
{\overline{E}^{\rm bath}_{\rm Hxc}(n)={E}_{\rm Hxc}(n)-{E}^{\rm imp}_{\rm
Hxc}(n)}.
\end{eqnarray}
This expression is convenient for developing approximate functionals.
For example, within the mean-field approximation, the Hxc functional for
the impurity is simplified as follows 
\begin{eqnarray}
{E}^{\rm imp}_{\rm
Hxc}(n)\rightarrow \dfrac{{U}}{4}{n}^2_{0},
\end{eqnarray}
thus leading, when combined with Eq.~(\ref{MF_approx_Hxc}), to
\begin{eqnarray}
\displaystyle \overline{E}^{\rm bath}_{\rm Hxc}(n)\rightarrow
\dfrac{{U}}{4}\sum_{i>0}{n}^2_{i}.
\end{eqnarray}
Better approximate functionals could be developed from the local density
approximation for a Luttinger liquid~\cite{PhysRevLett.90.146402}, in complete analogy with
the development of short-range functionals from the uniform electron
gas~\cite{toulda}. Work is currently in progress in this direction.\\

\begin{figure}
\caption{\label{fig:AC_graph_rep}
Graphical representation of the AC between the non-interacting (KS), the
embedded and the physical systems. Sites labelled with $\lambda U$ (in
red) and
$U$ (in blue) have partial and full on-site repulsion, respectively. Non-labelled sites (in
green) have no on-site repulsion. Local potentials (not represented)
are adjusted so that the sites occupation is constant along the AC. See text for further details.
}
\begin{center}
\begin{tabular}{c}
\hspace{1cm}\resizebox{11cm}{!}{\includegraphics{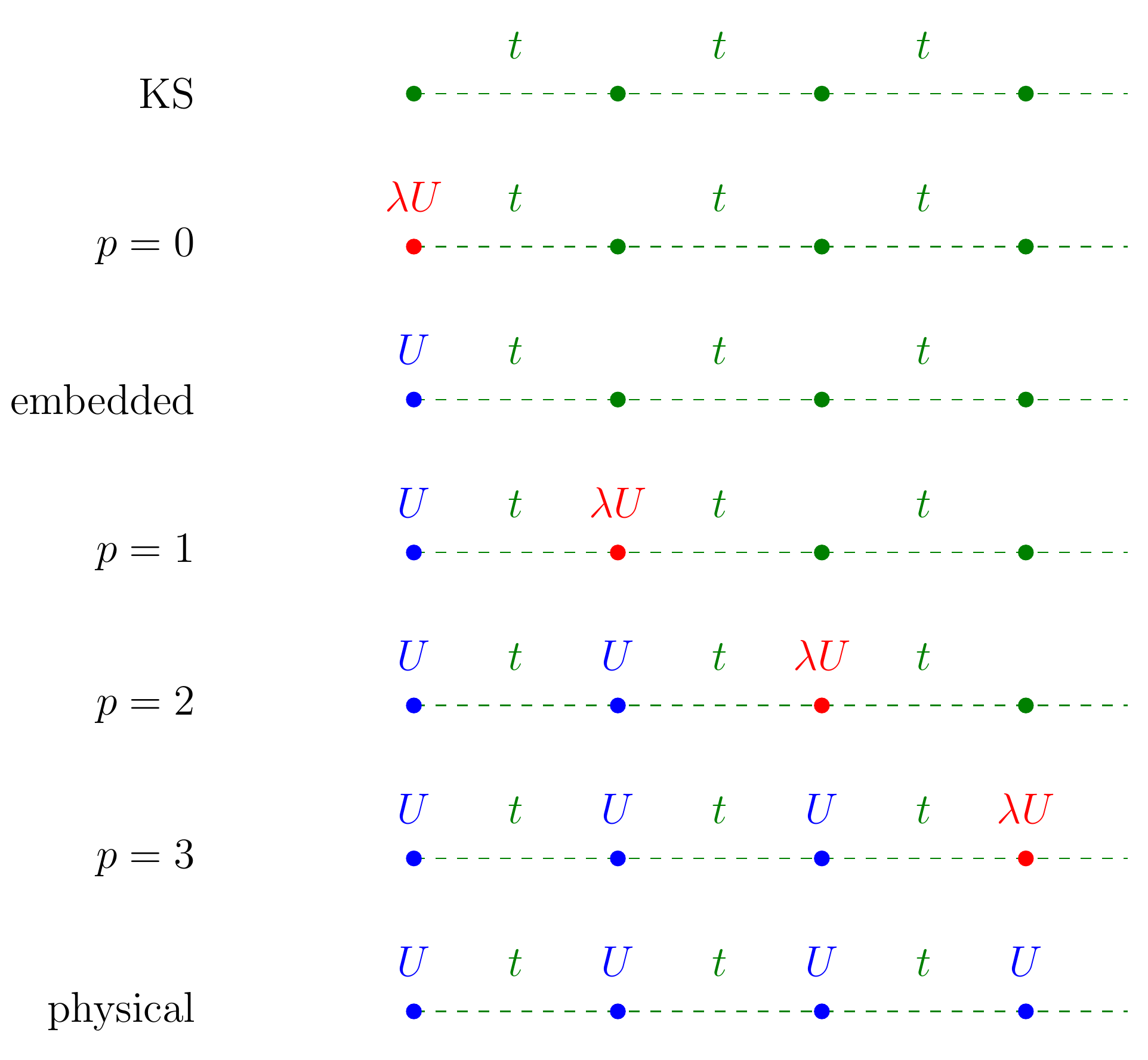}}
\end{tabular}
\end{center}
\end{figure}

Returning to the exact theory, a more explicit expression for the
complementary bath functional can be obtained within the AC formalism
when considering, in analogy with Eq.~(\ref{eq:AC_full_scale}), the following auxiliary
equations 
\begin{eqnarray}\label{eq:AC_impurities}
&&\Big(\hat{\mathcal{T}}+\hat{U}_{p-1}+\lambda{U}
\hat{n}_{p\uparrow}\hat{n}_{p\downarrow}+\sum_iv^{\lambda,p}_i\hat{n}_{i}\Big)\vert\Psi^{\lambda,p}\rangle
\nonumber
\\
&&=\mathcal{E}^{\lambda,p}\vert\Psi^{\lambda,p}\rangle,
\hspace{0.2cm} p=0, 1, \ldots,
\end{eqnarray}
where 
$\hat{U}_{p-1}=
{U}\sum^{p-1}_{i=0}
\hat{n}_{i\uparrow}\hat{n}_{i\downarrow}$ for $p\geq 1$, 
$\hat{U}_{-1}=0$, and $0\leq \lambda \leq1$. In such an AC the
on-site repulsion is switched on continuously site after site. A
graphical representation is given in Fig.~\ref{fig:AC_graph_rep}. 
The embedded impurity system is recovered when $p=0$ and $\lambda=1$ or,
equivalently, when $p=1$ and $\lambda=0$. From
Eqs.~(\ref{eq:LLimpfun}) and (\ref{eq:LL_imp_KSdecomp}), the site-occupation
constraint $n_{\Psi^{\lambda,p}}=n$ and the Hellmann--Feynman theorem, we
obtain   
\begin{eqnarray}\label{eq:impfun_AC_formula}
{E}^{\rm imp}_{\rm Hxc}(n)&=&
\int^1_0 \ddroit \lambda \dfrac{\ddroit
\mathcal{E}^{\lambda,0}}{\ddroit \lambda}+(v^{0,0}-v^{1,0}\vert n)
\nonumber\\
&=&U\int^1_0 \ddroit \lambda\;\langle \Psi^{\lambda,0}\vert
\hat{n}_{0\uparrow}\hat{n}_{0\downarrow}\vert\Psi^{\lambda,0}\rangle.
\end{eqnarray}
By introducing the multiple-impurity LL functional
\begin{eqnarray}
F^{\rm imp}_p(n)=
\underset{\Psi\rightarrow
n}{\rm
min}\left\{\langle\Psi\vert\hat{\mathcal{T}}+\hat{U}_p\vert\Psi\rangle\right\},
\end{eqnarray}
the (fully-interacting) Hxc functional becomes 
\begin{eqnarray}
{E}_{\rm Hxc}(n)=
\sum_{p\geq 0}F^{\rm imp}_{p}(n)-F^{\rm imp}_{p-1}(n),
\end{eqnarray}
since $F^{\rm imp}_{-1}(n)=\mathcal{T}_{\rm s}(n)$, thus leading to the
alternative expression (see
Eq.~(\ref{eq:Hxc_fun_ACexp})),
\begin{eqnarray}\label{eq:Hxc_fun_AC_alternative}
{E}_{\rm Hxc}(n)&=&
\sum_{p\geq 0}\Bigg(\int^1_0 \ddroit \lambda \dfrac{\ddroit
\mathcal{E}^{\lambda,p}}{\ddroit \lambda}+(v^{0,p}-v^{1,p}\vert n)
\Bigg)\nonumber\\
&=&
U\sum_{p\geq 0}\int^1_0 \ddroit \lambda\;\langle \Psi^{\lambda,p}\vert
\hat{n}_{p\uparrow}\hat{n}_{p\downarrow}\vert\Psi^{\lambda,p}\rangle.
\end{eqnarray}
Combining Eqs.~(\ref{eq:bathfun_as_diff}), (\ref{eq:impfun_AC_formula})
and (\ref{eq:Hxc_fun_AC_alternative}) finally gives the exact expression
\begin{eqnarray}\label{eq:AC_comp_bath_exp}
\overline{E}^{\rm bath}_{\rm Hxc}(n)=
U\sum_{p\geq 1}\int^1_0 \ddroit \lambda\;\langle \Psi^{\lambda,p}\vert
\hat{n}_{p\uparrow}\hat{n}_{p\downarrow}\vert\Psi^{\lambda,p}\rangle.
\end{eqnarray}    
If the exact complementary bath functional was known, its derivative could be computed
and, by solving Eq.~(\ref{eq:SC_eq_imp}) self-consistently, we would
obtain the exact embedded impurity wavefunction ${\Psi}^{\rm imp}$
that has the same sites occupation $n$ as the physical system
described by the Hubbard Hamiltonian in
Eq.~(\ref{eq:Hubb_hamil_ext_pot}):
\begin{eqnarray}\label{eq:nimp_equals_n}
n_{\Psi^{\rm imp}}=n.
\end{eqnarray}
If $n$ is known, the exact embedding potential
$v+\partial\overline{E}^{\rm bath}_{\rm Hxc}(n_{\Psi^{\rm
imp}})/\partial n$ can in fact be reached without knowing the
complementary bath functional. Indeed, according to the variational principle, for any
local potential $u\equiv\{u_i\}_i$, the embedded impurity wavefunction fulfils    
\begin{eqnarray}
\mathcal{E}^{\rm imp}(u)\leq
\langle {\Psi}^{\rm imp}\vert\hat{\mathcal{T}}+{U}\hat{n}_{0\uparrow}\hat{n}_{0\downarrow}
\vert {\Psi}^{\rm imp}\rangle
+(u\vert n_{\Psi^{\rm imp}}),
\end{eqnarray}
where $\mathcal{E}^{\rm imp}(u)$ is the ground-state energy of
$\hat{\mathcal{T}}+{U}\hat{n}_{0\uparrow}\hat{n}_{0\downarrow}+\sum_iu_i\hat{n}_{i}$.
Thus we obtain from Eqs.~(\ref{eq:LLimpfun}), (\ref{eq:SC_eq_imp}) and
(\ref{eq:nimp_equals_n}), the Legendre--Fenchel transform
expression~\cite{Eschrig,Kutzelnigg:2006,vanLeeuwen:2003,LFTransform-Lieb} 
\begin{eqnarray}
F^{\rm imp}(n)=
\underset{u
}{\rm sup}\Big\{
\mathcal{E}^{\rm imp}(u)-(u\vert n)
\Big\},
\end{eqnarray}
where $n$ is the input sites occupation and $u$ the embedding potential
to be optimized. In practice
approximations to $n$ could be used for example by performing a
mean-field description of the physical system. In this case, the
(approximate) embedding 
potential would enable to reproduce the mean-field occupation in both
the impurity and the bath.\\

Thus a connection with DMET
can be established (see Eq.~(5) in Ref.~\cite{PhysRevLett.109.186404}). A substantial
difference though is that, in DMET, the potential to be optimized
is the mean-field interaction operator $u$ that defines the mean-field
approximation to the Hubbard Hamiltonian, 
\begin{eqnarray}
\hat{H}^{\rm mf}(u)=\hat{\mathcal{T}}+u\sum_i\hat{n}_{i},
\end{eqnarray}
and the sites occupation 
to be reproduced by the ground state of
$\hat{H}^{\rm mf}(u)$ is the one obtained 
from the ground state of the DMET impurity Hamiltonian,
\begin{eqnarray}
\hat{H}^{\rm DMET}({u})=
\hat{\mathcal{T}}^{\rm DMET}
+U\hat{n}_{0\uparrow}\hat{n}_{0\downarrow}
+\tilde{u}\sum_{i>0}\hat{n}_i, 
\end{eqnarray} 
where 
$
\hat{\mathcal{T}}^{\rm DMET}=
\sum_{i>0,\sigma}{\rm \upsilon}(
\hat{a}^\dagger_{0\sigma}\hat{a}_{i\sigma}+
\hat{a}^\dagger_{i\sigma}\hat{a}_{0\sigma})
$. The terms $\rm \upsilon$ and $\tilde{u}$ are defined in
Ref.~\cite{PhysRevLett.109.186404} and depend
indirectly on $u$. 
Note that for simplicity DMET equations are written here for a uniform external potential.\\ 

At a given iteration $I$ of a DMET calculation, the mean-field potential is set to
$u_{I}$ and the impurity
Hamiltonian generates the ground-state sites occupation $n_{I}^{\rm imp}$. If
it exists a local potential $u_{\rm I+1}$ such that the ground state 
$
\Psi^{\rm
mf}(u_{I+1})
$
 of $\hat{H}^{\rm mf}(u_{I+1})$ has the same sites
occupation then, according to the variational principle, any trial
potential $u$ fulfils 
\begin{eqnarray}
\mathcal{E}^{\rm mf}(u)\leq \langle 
\Psi^{\rm
mf}(u_{I+1})
\vert\hat{\mathcal{T}}\vert 
\Psi^{\rm
mf}(u_{I+1})\rangle
+u(1\vert n_{I}^{\rm imp}), 
\end{eqnarray} 
where $\mathcal{E}^{\rm mf}(u)$ is the ground-state energy of
$\hat{H}^{\rm mf}(u)$, thus leading to the Legendre--Fenchel transform expression:
\begin{eqnarray}
\mathcal{T}_{\rm s}(n_{I}^{\rm imp})=\underset{u
}{\rm sup}\Big\{
\mathcal{E}^{\rm mf}(u)-u(1\vert n_{I}^{\rm imp})
\Big\}.
\end{eqnarray}
If a maximum exists, it then corresponds to $u_{I+1}$. The
updated occupation $n_{I+1}^{\rm imp}$ is then obtained from $u_{
I+1}$, thus generating a new potential until convergence is reached.
In this respect, DMET can be considered as a KS {\it Optimized Effective
Potential} (OEP) scheme 
since the mean-field  
Hamiltonian is in this context nothing but a non-interacting
Hamiltonian. The OEP is used to model the interaction on the impurity. A
formal analogy can actually be made with the range-separated KS-OEP
approach proposed in Ref.~\cite{pra_MBPTn-srdft}, where the long-range interaction is described with an OEP
while the short-range interaction is modeled with a density functional.
The OEP is then obtained from a density constraint, exactly
like in DMET.\\ 


Returning to SOET, let us finally mention that the theory can be extended to more impurity
sites simply by using the following partitioning of the LL functional,
\begin{eqnarray}\label{eq:FimpL_plus_Ebath}
F(n)=F^{\rm imp}_L(n)+\overline{E}^{\rm bath}_{{\rm Hxc},L}(n),
\end{eqnarray}  
where $L\geq 0$. In this case, the AC formula for the complementary bath functional
becomes 
\begin{eqnarray}\label{eq:AC_comp_bathL_exp}
\overline{E}^{\rm bath}_{{\rm Hxc},L}(n)=
U\sum_{p\geq L+1}\int^1_0 \ddroit \lambda\;\langle \Psi^{\lambda,p}\vert
\hat{n}_{p\uparrow}\hat{n}_{p\downarrow}\vert\Psi^{\lambda,p}\rangle.
\end{eqnarray}    

\section{Illustrative example: the two-site Hubbard
model}\label{sec:two_site_Hubb}

SOET is applied in this section to the simple two-site Hubbard
model with a uniform external potential. The analytical construction of the AC is presented
in Sec.~\ref{subsec:AC_for_two_sites}. The resulting integrand
expressions are then analyzed in
Sec.~\ref{subsec:results}.

\subsection{Symmetry breaking and restoration along the
AC}\label{subsec:AC_for_two_sites}

Let us consider a two-electron system described by the two-site Hubbard Hamiltonian
\begin{eqnarray}
\hat{H}&=&-t\sum_\sigma\Big(
\hat{a}^\dagger_{0\sigma}\hat{a}_{1\sigma}+
\hat{a}^\dagger_{1\sigma}\hat{a}_{0\sigma}
\Big)
+U_0\hat{n}_{0\uparrow}\hat{n}_{0\downarrow}
\nonumber\\
&&+U_1\hat{n}_{1\uparrow}\hat{n}_{1\downarrow}
+v_0\hat{n}_{0}+v_1\hat{n}_{1}.
\end{eqnarray}
Since we are interested here in the
singlet ground state only, the matrix representation of the Hamiltonian can
be reduced to the basis of the "doubly-occupied site" states $\vert
D_i\rangle=\hat{a}^\dagger_{i\uparrow}\hat{a}^\dagger_{i\downarrow}\vert{\rm
vac}\rangle$ with $i=0$ or $1$, and $\vert
S\rangle=1/\sqrt{2}(\hat{a}^\dagger_{0\uparrow}\hat{a}^\dagger_{1\downarrow}-\hat{a}^\dagger_{0\downarrow}\hat{a}^\dagger_{1\uparrow})\vert
{\rm vac}\rangle$ that corresponds to singly-occupied sites, thus leading to       
\begin{eqnarray}\label{Hmatrix}
\left[ \hat{H} \right]
&=&
\left[
\begin{array}{c c c}
U_0+v_0-v_1 & 0 & -\sqrt{2}t\\
0 & U_1+v_1-v_0 & -\sqrt{2}t\\ 
-\sqrt{2}t & -\sqrt{2}t & 0 
\end{array}
\right]
+(v_0+v_1) 
.
\end{eqnarray}
For simplicity we choose for the physical Hamiltonian $v_0=v_1=0$ and
$U_0=U_1=U$. Consequently the site occupation is uniform: 
\begin{eqnarray}\label{eq:SO_constraint_two_sites}
n_0=n_1=1.
\end{eqnarray} 
We obtain by diagonalization the well-known expressions for the ground-state energy 
\begin{eqnarray}\label{eq:Hubb_GS_ener}
E(U)=\dfrac{1}{2}\Big(U-\sqrt{U^2+16t^2}\Big),
\end{eqnarray}
and the corresponding wavefunction
\begin{eqnarray}
\vert \Psi(U)\rangle=d(U)\Big(\vert D_0\rangle+\vert
D_1\rangle\Big)+\sqrt{1-2d^2(U)}\vert S\rangle,
\end{eqnarray}
where
\begin{eqnarray}
d^2(U)=\dfrac{E^2(U)}{8t^2+2E^2(U)}
=\dfrac{E(U)}{4E(U)-2U},
\end{eqnarray}
since $E^2(U)-UE(U)-4t^2=0$, thus leading to the more explicit expression
\begin{eqnarray}\label{eq:explicit_exp_2bleocc}
d^2(U)=
\dfrac{1}{4}\left(1
-\dfrac{1}{\sqrt{1+\dfrac{16}{(U/t)^2}}}
\right)
.
\end{eqnarray}
Note the Hellmann--Feynman theorem,
\begin{eqnarray}\label{Hell--Feyn_th_2sites}
d^2(U)=\dfrac{1}{2}\dfrac{\ddroit E(U)}{\ddroit U},
\end{eqnarray}
that will be used in the following.
The double occupancy is the same for both
sites and equal to
\begin{eqnarray}\label{eq:same_2bleocc}
\langle \Psi(U)
\vert\hat{n}_{0\uparrow}\hat{n}_{0\downarrow}\vert
\Psi(U)\rangle
&=&\langle \Psi(U)
\vert\hat{n}_{1\uparrow}\hat{n}_{1\downarrow}\vert
\Psi(U)\rangle
=
d^2(U).
\end{eqnarray}
Along the AC described in Eq.~(\ref{eq:AC_impurities}) and
Fig.~\ref{fig:AC_graph_rep}, symmetry in the on-site repulsions is
broken in the auxiliary Hamiltonian operator. This is clear for the embbeded
impurity where $U_0=U$ and $U_1=0$. Nevertheless, symmetry can be
restored in the Hamiltonian matrix simply by adjusting the local
potential components $v_0$ and $v_1$ such that 
\begin{eqnarray}\label{eq:emb_pot_two_sites_Hubb0}
U_0+v_0-v_1=U_1+v_1-v_0,
\end{eqnarray}
thus leading to
\begin{eqnarray}\label{eq:emb_pot_two_sites_Hubb}
v_1-v_0=\dfrac{U_0-U_1}{2}.
\end{eqnarray}
The latter condition defines uniquely (up to a constant) the local
embedding potential that gives a uniform site occupation. In this case the effective
on-site repulsion is the same on each site and is simply expressed as
\begin{eqnarray}
U_{\rm eff}=\dfrac{U_0+U_1}{2}.
\end{eqnarray}
Therefore the AC can be constructed analytically as follows for the
impurity ($p=0$),
\begin{eqnarray}\label{eq:auxHamil_AC_bath}
U_0=\lambda U,\, U_1=0 &\,\rightarrow\,&
v_0^{\lambda,0}=0,\,v_1^{\lambda,0}={\lambda U}/{2},\,
U_{\rm eff}=\lambda U/2,
\end{eqnarray}    
and for the bath ($p=1$),
\begin{eqnarray}\label{eq:AC_p1}
U_0=U,\, U_1=\lambda U &\,\rightarrow\,&
v_0^{\lambda,1}=0,\,v_1^{\lambda,1}={(1-\lambda)U}/{2},\,
U_{\rm eff}=(1+\lambda) U/2
.
\end{eqnarray}
Consequently the wavefunctions along the AC for the
bath and the impurity are 
\begin{eqnarray}
\Psi^{\lambda,0}=\Psi\big(\lambda U/2\big),
\end{eqnarray}
and
\begin{eqnarray}\label{eq:wf_AC_bath}
\Psi^{\lambda,1}=\Psi\big((1+\lambda) U/2\big),
\end{eqnarray}
respectively. The corresponding double occupancies, that are nothing but
Hxc integrands per unit of $U$ for the impurity and the bath (see
Eqs.~(\ref{eq:impfun_AC_formula}) and (\ref{eq:AC_comp_bath_exp})), respectively, can then be
expressed, according to Eqs.~(\ref{eq:explicit_exp_2bleocc}) and
(\ref{eq:same_2bleocc}), as   
\begin{eqnarray}\label{eq:2ble_occ_imp_explicit_exp}
\langle \Psi^{\lambda,0}
\vert\hat{n}_{0\uparrow}\hat{n}_{0\downarrow}\vert
\Psi^{\lambda,0}\rangle=
\dfrac{1}{4}\left(1
-\dfrac{1}{\sqrt{1+\dfrac{64}{\lambda^2(U/t)^2}}}
\right),
\end{eqnarray} 
and
\begin{eqnarray}\label{eq:2ble_occ_bath_explicit_exp}
\langle \Psi^{\lambda,1}
\vert\hat{n}_{1\uparrow}\hat{n}_{1\downarrow}\vert
\Psi^{\lambda,1}\rangle
=
\dfrac{1}{4}\left(1
-\dfrac{1}{\sqrt{1+\dfrac{64}{(1+\lambda)^2(U/t)^2}}}
\right).
\end{eqnarray} 
Note that the embedded impurity wavefunction is obtained when
$\lambda=1$ along the AC for the impurity, and $\lambda=0$ along the AC
for the bath:
\begin{eqnarray}
\Psi^{\rm imp}= \Psi^{1,0}=\Psi^{0,1}=\Psi(U/2).
\end{eqnarray}
Since, according to Eqs.~(\ref{Hmatrix}) and
(\ref{eq:auxHamil_AC_bath}), 
\begin{eqnarray}
&&\langle {\Psi}^{\rm imp}\vert\hat{\mathcal{T}}+{U}\hat{n}_{0\uparrow}\hat{n}_{0\downarrow}
+(U/2)\hat{n}_1\vert {\Psi}^{\rm imp}\rangle
=E(U/2)
+U/2,
\end{eqnarray}
it comes from the site occupation constraint in
Eq.~(\ref{eq:SO_constraint_two_sites}) that
\begin{eqnarray}
\langle {\Psi}^{\rm imp}\vert\hat{\mathcal{T}}+{U}\hat{n}_{0\uparrow}\hat{n}_{0\downarrow}
\vert {\Psi}^{\rm imp}\rangle
=E(U/2)
.
\end{eqnarray}
Moreover, according to Eqs.~(\ref{Hell--Feyn_th_2sites}) and
(\ref{eq:wf_AC_bath}), 
the double occupancy can be rewritten along the AC for the bath as 
\begin{eqnarray}\label{eq:2ble_occ_AC_bath}
\langle \Psi^{\lambda,1}
\vert\hat{n}_{1\uparrow}\hat{n}_{1\downarrow}\vert
\Psi^{\lambda,1}\rangle
=\dfrac{1}{U}\dfrac{\ddroit E\Big((1+\lambda)U/2)\Big)}{\ddroit
\lambda},
\end{eqnarray} 
thus giving with Eq.~(\ref{eq:AC_comp_bath_exp}), 
\begin{eqnarray}
\overline{E}^{\rm bath}_{\rm Hxc}(n)=E(U)-E(U/2),
\end{eqnarray}
or, more explicitly,
\begin{eqnarray}
\overline{E}^{\rm bath}_{\rm Hxc}(n)/U&=&
\dfrac{1}{4}
-\dfrac{3/4}{
\sqrt{1+\dfrac{64}{(U/t)^2}}
+
2\sqrt{1+\dfrac{16}{(U/t)^2}}
}
.
\end{eqnarray}
As expected, the physical energy $E(U)$ is recovered when adding
contributions from the impurity and the bath:
\begin{eqnarray}
E(U)=
\langle {\Psi}^{\rm imp}\vert\hat{\mathcal{T}}+{U}\hat{n}_{0\uparrow}\hat{n}_{0\downarrow}
\vert {\Psi}^{\rm imp}\rangle
+\overline{E}^{\rm bath}_{\rm Hxc}(n)
.
\end{eqnarray}

Let us finally mention that, in the conventional AC (see
Eq.~(\ref{eq:AC_full_scale})),
the on-site repulsion is scaled on both the
impurity and the bath sites. In this case the interaction on the impurity
is not separated from the repulsions in the bath. The corresponding AC will
therefore be defined as
\begin{eqnarray}\label{eq:auxHamil_convAC}
U_0=U_1=\lambda U &\,\rightarrow\,&
v_0^{\lambda}=v_1^{\lambda}=0,\,
U_{\rm eff}=\lambda U,
\end{eqnarray}    
and
\begin{eqnarray}
\Psi^\lambda=\Psi(\lambda U).
\end{eqnarray}
As a result the AC integrand can be expressed as
\begin{eqnarray}
\langle \Psi^\lambda \vert \hat{U}\vert
\Psi^\lambda\rangle=2Ud^2(\lambda U),
\end{eqnarray}
thus leading to the explicit expression 
\begin{eqnarray}\label{eq:conv_AC_int_explicit_exp}
\dfrac{
\langle \Psi^\lambda \vert \hat{U}\vert
\Psi^\lambda\rangle
}{U}=
\dfrac{1}{2}\left(1
-\dfrac{1}{\sqrt{1+\dfrac{16}{\lambda^2(U/t)^2}}}
\right).
\end{eqnarray}
Note that, according to Eq.~(\ref{Hell--Feyn_th_2sites}), the Hxc integrand can
also be rewritten as  
\begin{eqnarray}\label{eq:convAC_HellTh_int}
\langle \Psi^\lambda \vert \hat{U}\vert
\Psi^\lambda\rangle=
\dfrac{\ddroit E\big(\lambda U\big)}{\ddroit\lambda}.
\end{eqnarray}
Finally by rewriting the double
occupancy along the AC for the impurity (see
Eq.~(\ref{eq:2ble_occ_imp_explicit_exp})) as follows,
\begin{eqnarray}
\langle \Psi^{\lambda,0}
\vert\hat{n}_{0\uparrow}\hat{n}_{0\downarrow}\vert
\Psi^{\lambda,0}\rangle=\dfrac{1}{U}
\dfrac{\ddroit E\big(\lambda U/2\big)}{\ddroit\lambda},
\end{eqnarray}  
we see from
Eqs.~(\ref{eq:Hxc_fun_ACexp}), (\ref{eq:Hxc_fun_AC_alternative}),
(\ref{eq:2ble_occ_AC_bath}) and (\ref{eq:convAC_HellTh_int}) that, as
expected, 
both the conventional Hxc integrand
and the sum of bath and impurity integrands lead to the same Hxc energy after
integration over [0,1]:
\begin{eqnarray}
&&{E}_{\rm Hxc}(n)=E(U)-E(0)
=\Big(E(U/2)-E(0)\Big)+
\Big(E(U)-E(U/2)\Big),
\end{eqnarray}   
or, more explicitly,
\begin{eqnarray}
&&{E}_{\rm Hxc}(n)/U=
\dfrac{1}{2}\Bigg(1+\dfrac{4}{U/t}
-\sqrt{1+\dfrac{16}{(U/t)^2}}
\Bigg).
\end{eqnarray}

\subsection{Results and discussion}\label{subsec:results}
\begin{figure}
\caption{\label{fig:2site_Hubb_imp_AC} 
Double occupancy along the AC for the impurity plotted with respect to
the interaction (top panel) and correlation (bottom panel) strenghts. See text for
further details. 
}
\begin{center}
\begin{tabular}{c}
\hspace{1cm}\resizebox{11cm}{!}{\includegraphics{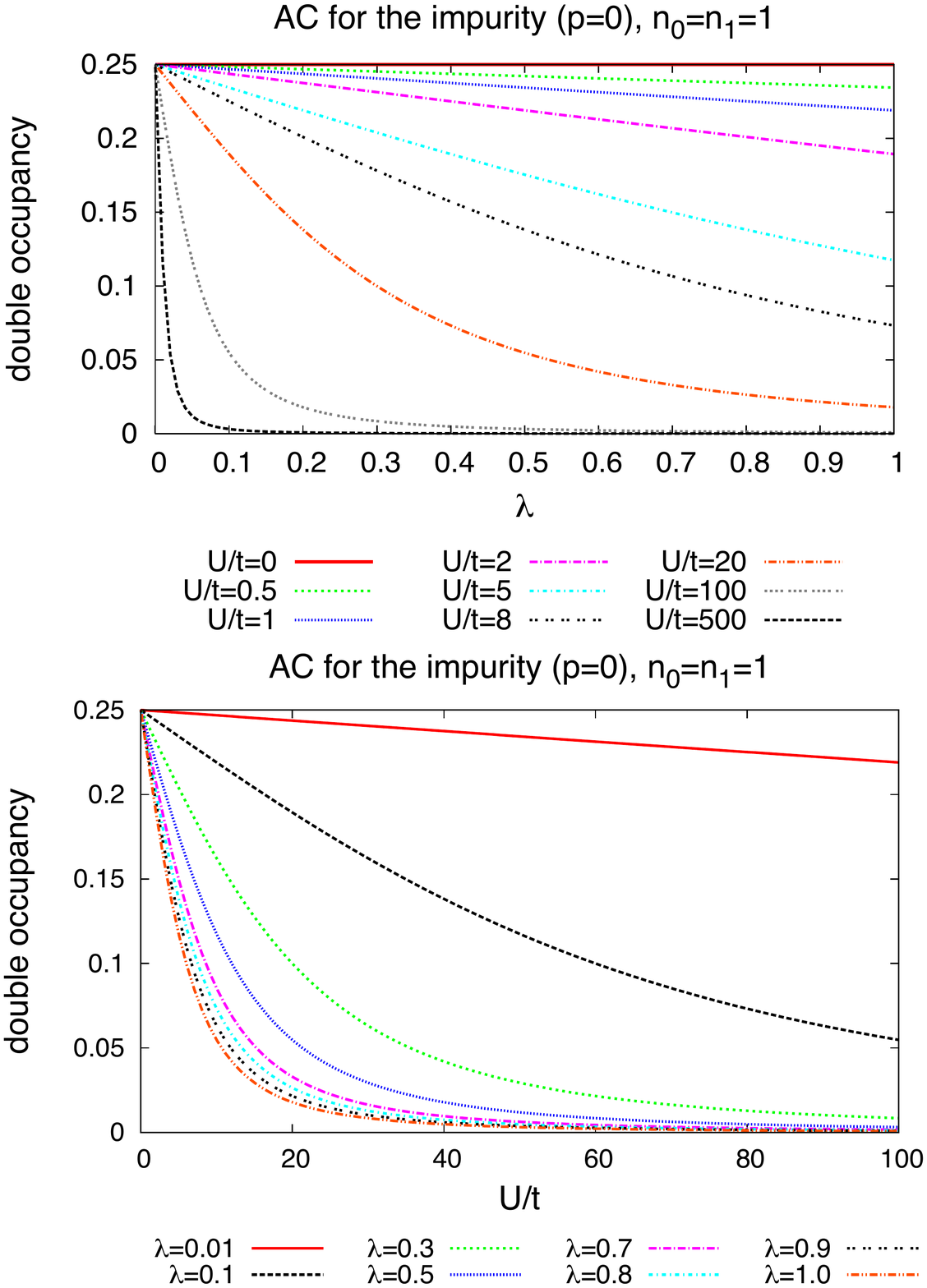}}
\end{tabular}


\end{center}
\end{figure}

\begin{figure}
\caption{\label{fig:2site_Hubb_bath_AC}
Double occupancy along the AC for the bath plotted with respect to
the interaction (top panel) and correlation (bottom panel) strenghts. See text for
further details. 
}
\begin{center}
\begin{tabular}{c}
\hspace{1cm}\resizebox{11cm}{!}{\includegraphics{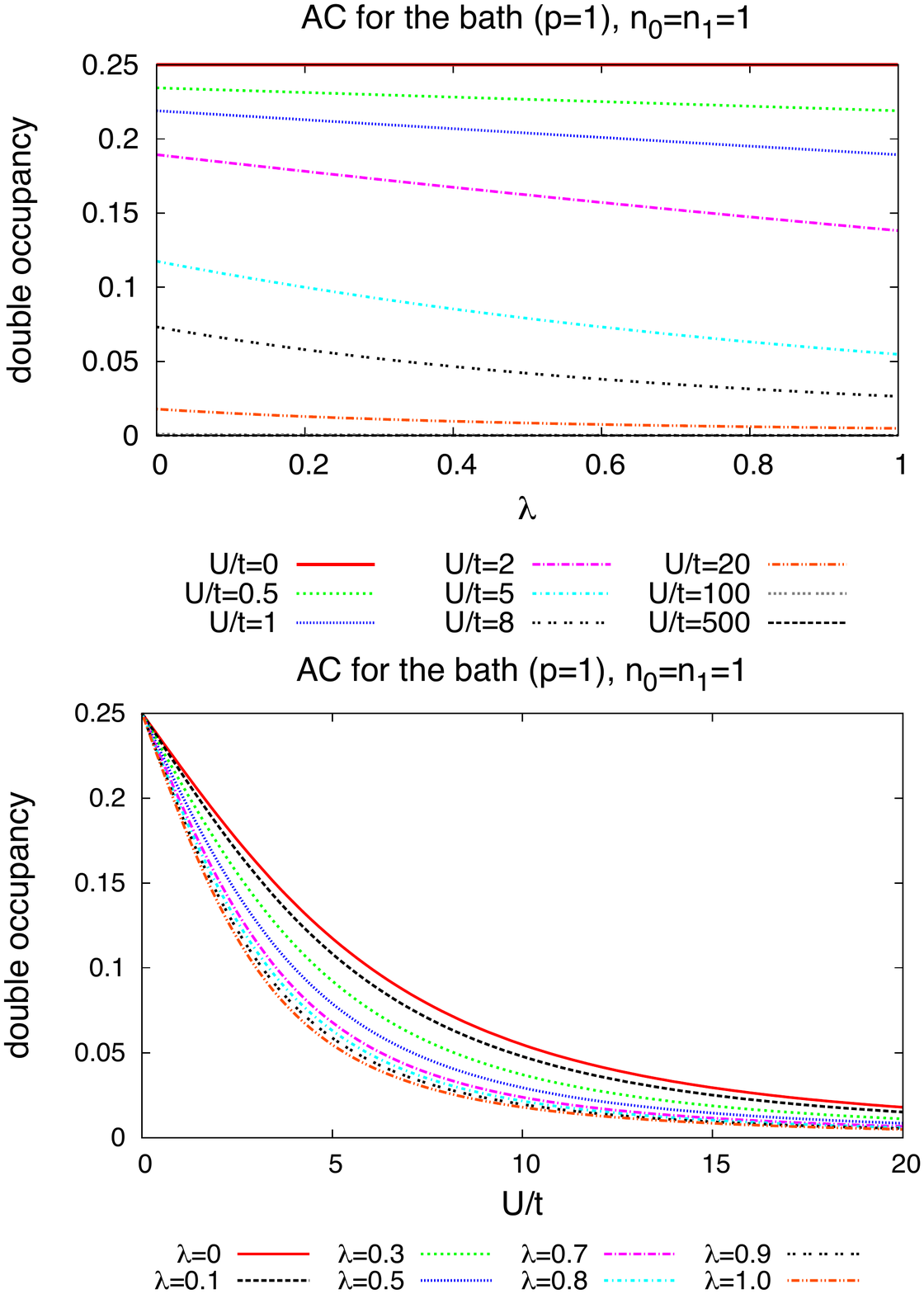}}
\end{tabular}


\end{center}
\end{figure}
\begin{figure}
\caption{\label{fig:2site_Hubb_bath_comparACs}
Hxc integrands per unit of U plotted for the bath, bath+impurity and the
two sites. Various correlation regimes are considered: $U/t=0.05$ (red), $0.2$
(green), 1 (grey), 10 (black), 30 (blue) and 300 (purple). See text for
further details. 
}
\begin{center}
\begin{tabular}{c}
\resizebox{15cm}{!}{\includegraphics{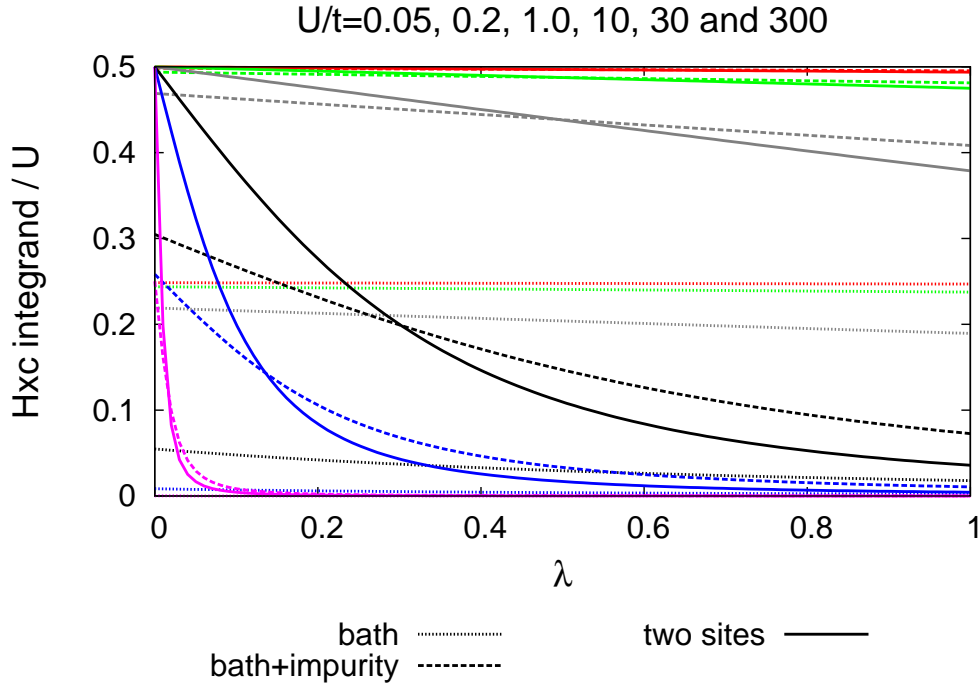}}
\end{tabular}
\end{center}
\end{figure}

Double occupancies along the AC for the impurity and the bath (see
Eqs.~(\ref{eq:2ble_occ_imp_explicit_exp}) and
(\ref{eq:2ble_occ_bath_explicit_exp})), that can
be identified as Hxc integrands per unit of $U$, have been plotted for
various correlation regimes. Results are shown in
Figs.~\ref{fig:2site_Hubb_imp_AC} and \ref{fig:2site_Hubb_bath_AC}. For
analysis purposes, comparison is made between the sum of these two
integrands and the conventional Hxc integrand in
Eq.~(\ref{eq:conv_AC_int_explicit_exp}). 
Results are shown in
Fig.~\ref{fig:2site_Hubb_bath_comparACs}.\\ 

Let us first focus on the impurity. As expected from accurate quantum
chemical calculations of the AC for the H$_2$ molecule at
equilibrium~\cite{AMTOptseplrsr}, the integrand varies linearly with the interaction
strength $\lambda$ in the weakly correlated regime ($U/t<<1$). 
The integrand gains curvature as $U/t$ increases. In the strongly
correlated regime ($U/t>>1$), the double occupancy of the impurity becomes
zero for large interaction strengths (see the bottom panel in
Fig.~\ref{fig:2site_Hubb_imp_AC}). It only varies in the vicinity of the
non-interacting case ($\lambda=0$) with a large negative slope. This
pattern was found by Teale~\etal~for the stretched H$_2$ molecule
when computing the AC at the {\it Full Configuration Interaction} (FCI) level in a large basis set~(see Fig.~5
(a) in Ref.~\cite{AMTOptseplrsr}). One
should mention that, along the AC for the impurity, the on-site repulsion in the bath is set to zero. Comparison with
Ref.~\cite{AMTOptseplrsr} is then more relevant when considering the conventional AC
integrand for the two sites. As shown in
Fig.~\ref{fig:2site_Hubb_bath_comparACs}, all correlation regimes
observed numerically for H$_2$ along the AC and the bond breaking coordinate are qualitatively well
reproduced by applying SOFT to the two-site Hubbard Hamiltonian.\\

Turning to the AC for the bath, we first notice in the top panel of
Fig.~\ref{fig:2site_Hubb_bath_AC} that the double occupancy of the bath
remains essentially linear in $\lambda$ for all correlation regimes, in
contrast to the AC for the impurity. This is due to the fact that
the on-site repulsion is already switched on for the impurity. Scaling
is only used for the repulsion on the bath site so that the embedded impurity
($\lambda=0$) can be connected with the physical system 
($\lambda=1$). As expected, the two systems exhibit different double
occupancies even though they have the same sites occupation (see the
bottom panel in Fig.~\ref{fig:2site_Hubb_bath_AC}). Variations with
the correlation strength are consistent with those obtained for the
1D-Hubbard model (see case $<n>=1$ in Fig.~2 of
Ref.~\cite{PhysRevLett.109.186404}).\\                

Let us finally discuss the separation of interactions between the impurity
and the bath. As mentioned previously and illustrated in
Fig.~\ref{fig:2site_Hubb_bath_comparACs}, both the conventional and
bath+impurity AC integrands give the same total Hxc energy after
integration over $[0,1]$ but they of course differ along the AC. In the
weakly correlated regime, both the bath and the impurity contribute
significantly to the Hxc energy. Note that, in practical calculations,
we want to describe the embedded impurity only, meaning that the
contributions of the bath to the Hxc energy will be described by a
site-occupation functional. It becomes clear from this simple example that
the latter can contribute significantly and that approximations beyong
the mean field (which is exact only when $U/t=0$) will be needed.
Returning to the AC, in the strongly correlated regime, the contribution
of the bath
becomes negligible for all interaction strengths. 
Note that, for  
$\lambda=0$, the impurity and conventional Hxc integrands are equal to
$1/4$ and $1/2$, respectively. Since both are non-zero only for small
interaction strengths and behave similarly (with a large negative
slope), they both integrate to the same Hxc energy simply because the effective on-site
repulsion along the AC is smaller for the impurity alone ($U_{\rm eff}=\lambda U/2$)
than for the two sites ($U_{\rm eff}=\lambda U$).

\section{Applying SOFT to quantum
chemistry}\label{sec:SOFT_in_QC}

We propose in this section to apply SOFT to quantum
chemical Hamiltonians. The formulation of a KS scheme in this context
is discussed in Sec.~\ref{subsec:ks-soft_QC}. We then show in
Sec.~\ref{subsec:cassoft} that SOET can lead to an alternative CASDFT
method.   
\subsection{
KS-SOFT for a quantum chemical Hamiltonian}\label{subsec:ks-soft_QC}

In order to adapt SOFT to quantum chemistry, let us first consider the
orthonormal basis of molecular orbitals $\{\phi_{p}(\mathbf{r})\}_p$
obtained for non-interacting electrons, so that
\begin{eqnarray}\label{eq:orbitals_soft_QC}
\hat{h}\phi_{p}=h_{pp}\phi_{p},
\end{eqnarray}
with $
\hat{h}=-\dfrac{1}{2}\nabla_{\mathbf{r}}^2+v_{\rm
ne}(\mathbf{r}).
$
This is obviously not a good starting point for conventional
quantum chemical calculations but it is convenient for
deriving a KS-SOFT scheme in this context. As explained further in the
following, the theory can in principle be adapted to different choices
of orbitals. Using Eq.~(\ref{eq:orbitals_soft_QC}) leads to the
following second-quantized expression for the molecular Hamiltonian
$\hat{H}=\hat{T}+\hat{W}_{\rm ee}+\hat{V}_{\rm ne}$,
\begin{eqnarray}
\hat{H}&=&
\sum_ph_{pp}\hat{n}_p+\hat{U}^{\rm HF}+
\hat{\mathcal{W}}_{\rm
ee},
\end{eqnarray}
where $\hat{\mathcal{W}}_{\rm ee}=
\hat{W}_{\rm
ee}
-\hat{U}^{\rm HF}$ denotes the fluctuation potential. The Hartree--Fock
(HF) potential operator can be decomposed as follows,
\begin{eqnarray}
&&\hat{U}^{\rm
HF}=\sum_pu_{pp}\hat{n}_{p}+\hat{\mathcal{T}}^{\rm HF},\label{eq:UHFdecomp}
\\
&&\hat{\mathcal{T}}^{\rm HF}=\sum_{p\neq
q,\sigma}u_{pq}\hat{a}_{p\sigma}^\dagger\hat{a}_{q\sigma},\label{eq2}\\
&&u_{pq}=
\sum_{r\in\rm occ} 2
\langle pr\vert qr\rangle
-
\langle pr\vert rq\rangle,\label{eq:HFpot_occ}
\end{eqnarray}
where $\langle pq\vert rs\rangle$ denotes a regular two-electron
integral. The first term on the right-hand side of Eq.~(\ref{eq:UHFdecomp})
is local in the orbital space while the second one is non-local and thus
enables hopping between occupied and unoccupied orbitals, hence the
notation $\hat{\mathcal{T}}^{\rm HF}$ in analogy with the kinetic energy operator in the Hubbard
Hamiltonian. Therefore the molecular Hamiltonian becomes 
\begin{eqnarray}\label{eq:mol_hamil_like_Hubb}
\hat{H}&=&
\hat{\mathcal{T}}^{\rm HF}+\hat{\mathcal{W}}_{\rm
ee}+\sum_p\varepsilon_p\hat{n}_p,
\end{eqnarray}
with $\varepsilon_p=h_{pp}+u_{pp}$. This expression enables a direct
comparison with the Hubbard Hamiltonian in
Eq.~(\ref{eq:Hubb_hamil_ext_pot}): the sites are now molecular orbitals
and their energies $\varepsilon\equiv\{\varepsilon_p\}_p$ play the role
of the external potential.\\

Following Sec.~\ref{subsec:ks-soft}, we introduce the LL functional      
\begin{eqnarray}
F(n)=
\underset{\Psi\rightarrow
n}{\rm
min}\left\{\langle\Psi\vert\hat{\mathcal{T}}^{\rm HF}+\hat{\mathcal{W}}_{\rm ee}\vert\Psi\rangle\right\},
\end{eqnarray}
so that the exact ground-state energy, that is the FCI energy in this
context, can be written as 
\begin{eqnarray}\label{eq:FCIener_vp_n}
\displaystyle E({\varepsilon})=\underset{n}{\rm min}\Big\{F(n)+
({\varepsilon}\vert n)\Big\}.
\end{eqnarray}
Let us consider the KS partitioning, 
\begin{eqnarray}\label{eq:HxcSOQCfun_def}
F(n)=\mathcal{T}^{\rm HF}_{\rm s}(n)+E_{\rm c}(n),
\end{eqnarray}
where $
\mathcal{T}^{\rm HF}_{\rm s}(n)=
\underset{\Psi\rightarrow
n}{\rm
min}\left\{\langle\Psi\vert\hat{\mathcal{T}}^{\rm HF}\vert\Psi\rangle\right\}
$ and, in contrast to Eq.~(\ref{eq:HxcSOfun_def}), no
Hartree and exchange energy contributions have been introduced. Indeed, the fluctuation
potential induces correlation effects only since, when written with normal
ordering~\cite{lindgren}, it generates double excitations only.
Consequently, within
KS-SOFT, the FCI
energy is obtained variationally as follows, 
\begin{eqnarray}\label{eq:FCIener_var_soft} 
E({\varepsilon})=
\underset{\Psi
}{\rm
min}\left\{\langle\Psi\vert\hat{\mathcal{T}}^{\rm HF}
\vert\Psi\rangle+
{E}_{\rm c}(n_\Psi)
+({\varepsilon}\vert n_\Psi)\right\},
\end{eqnarray}
where the minimizing wavefunction $\Psi^{\rm KS}$
fulfils the self-consistent
equation
\begin{eqnarray}\label{eq:ks-soft_eq_QC}
\Bigg(\hat{\mathcal{T}}^{\rm HF}+
\sum_p\Bigg[{\varepsilon_p}+\dfrac{\partial {E}_{\rm
c}(n_{{\Psi^{\rm KS}}
})}
{\partial
n_p}\Bigg]\hat{n}_p\Bigg)\vert{\Psi^{\rm KS}}\rangle
={\mathcal{E}^{\rm
KS}}\vert{\Psi^{\rm KS}
}\rangle.
\end{eqnarray}
Eqs.~(\ref{eq:FCIener_var_soft}) and (\ref{eq:ks-soft_eq_QC}) should in principle enable to
recover the FCI energy without treating electron
correlation explicitly. Orbital rotations would therefore be sufficient.
One should of course investigative potential $\varepsilon$-{representability} problems for example by computing Legendre--Fenchel
transforms. Approximate functionals may also be developed from model
Hamiltonians such as the Hubbard Hamiltonian. 
Let us finally note that different orbitals could be used in
Eq.~(\ref{eq:ks-soft_eq_QC}). One would need to adapt the correlation
functional in order to recover the correct FCI energy. Work is in
progress in these directions. 

\subsection{CASSOFT approach}\label{subsec:cassoft}

\begin{figure}
\caption{\label{fig:active_space_graph_rep} 
Graphical representation of the active spaces ${\mathcal A}_{r,s}$.
Dashed blue lines represent active orbitals in ${\mathcal A}_{r_0,s_0}$. Full green lines are either
inactive or virtual orbitals. See text for further details. 
}
\begin{center}
\begin{tabular}{c}
\hspace{1cm}\resizebox{9cm}{!}{\includegraphics{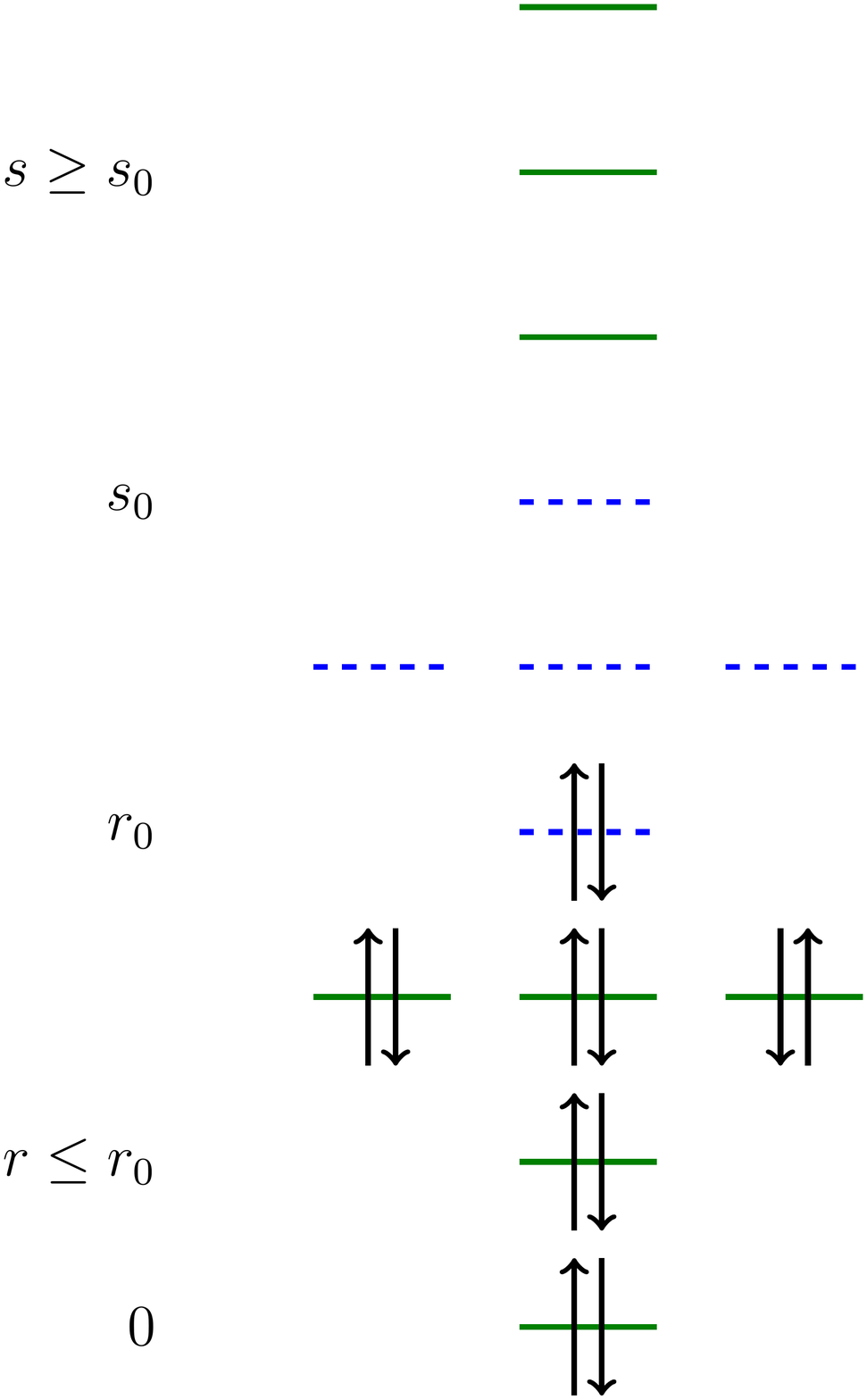}}
\end{tabular}
\end{center}
\end{figure}

Following Knizia and Chan~\cite{JCTC13_Chan_DMET_QC}, 
we propose in this section to apply SOET 
to the molecular Hamiltonian in Eq.~(\ref{eq:mol_hamil_like_Hubb}). The
impurity sites will be the
active orbitals ($u,v,\ldots$) whose selection is usually based on chemical
intuition, while the inactive and
virtual orbitals correspond to the bath. A graphical representation is
given in Fig.~\ref{fig:active_space_graph_rep}. As already mentioned for SOET, the
separation of correlation effects is not controlled here by a single
continuous parameter like in range-separated DFT. In the spirit of a
regular CASSCF calculation, it rather relies on
the selection of active orbitals. This ensures that only correlation effects
within the active space are described explicitly in WFT while the
remaining correlation, including core correlation, is 
modeled by an orbital-occupation functional.  
Separating correlation effects in the active orbital space
$\mathcal{A}$ from the remaining ones leads to the alternative partitioning of the LL
functional, 
\begin{eqnarray}\label{eq:part_F_imp_dyn}
F(n)=F^{\rm imp}_{ \mathcal{A}}(n)+\overline{E}^{\rm bath}_{\rm c}(n),
\end{eqnarray}
where 
\begin{eqnarray}\label{eq:Fimp_LL_qc}
F^{\rm imp}_{ \mathcal{A}}(n)
=\underset{\Psi\rightarrow
n}{\rm
min}\left\{
\langle\Psi\vert\hat{\mathcal{T}}^{\rm HF}+
\hat{\mathcal{W}}_{\rm \mathcal{A}}
\vert\Psi\rangle
\right\},
\end{eqnarray}
and $\hat{\mathcal{W}}_{\rm \mathcal{A}}$ is the
reduction of the fluctuation potential operator to the active orbital space: 
\begin{eqnarray}\label{eq:fluctu_pot_active}
\hat{\mathcal{W}}_{\rm \mathcal{A}}
&=&{\dfrac{1}{2}\sum_{u,v,x,y,\sigma,\sigma'}\langle uv\vert 
xy
\rangle\;
\hat{a}^\dagger_{u\sigma}\hat{a}^\dagger_{v\sigma'}\hat{a}_{y\sigma'}\hat{a}_{x\sigma}
}
\nonumber\\
&&-\sum_{u,v}
\sum_{x\in\rm occ} \Big(2
\langle ux\vert vx\rangle
-
\langle ux\vert xv\rangle\Big)
\hat{a}_{u\sigma}^\dagger\hat{a}_{v\sigma}.
\end{eqnarray}
Note that the HF potential in the second term on the right-hand side of
Eq.~(\ref{eq:fluctu_pot_active}) is here calculated with the active
orbitals that are doubly occupied, by analogy with Eq.~(31) in Ref.~\cite{JCP11_Chan_DMFT_for_QC}. Combining Eqs.~(\ref{eq:FCIener_vp_n}),
(\ref{eq:part_F_imp_dyn}) and (\ref{eq:Fimp_LL_qc}) leads to the exact
variational expression 
\begin{eqnarray}
E({\varepsilon})=
\underset{\Psi
}{\rm
min}\Big\{
\langle\Psi\vert\hat{\mathcal{T}}^{\rm HF}+\hat{\mathcal W}_{\mathcal A}
\vert\Psi\rangle+
\overline{E}^{\rm bath}_{\rm c}(n_\Psi)
+({\varepsilon}\vert n_\Psi)\Big\},
\end{eqnarray}
where the minimizing wavefunction $\Psi_{\mathcal A}^{\rm imp}$ fulfils the
self-consistent equation
\begin{eqnarray}\label{eq:sc_cassoft}
\Bigg(
\hat{\mathcal{T}}^{\rm HF}+\hat{\mathcal W}_{\mathcal A}
+\sum_p\Bigg[{\varepsilon_p}+\dfrac{\partial
\overline{E}^{\rm bath}_{\rm c}
(n_{{\Psi_{\mathcal A}^{\rm imp}}
})}
{\partial
n_p}\Bigg]\hat{n}_p\Bigg)\vert{\Psi_{\mathcal A}^{\rm imp}}\rangle
={\mathcal{E}
_{\mathcal A}^{\rm imp}}\vert{\Psi_{\mathcal A}^{\rm imp}
}\rangle.
\end{eqnarray}
The method will be referred to as CASSOFT. Interestingly the impurity Hamiltonian is very similar to the one
proposed by Zgid and Chan in Ref.~\cite{JCP11_Chan_DMFT_for_QC}. A major difference is that
the embedding potential $\varepsilon+\partial\overline{E}^{\rm bath}_{\rm
c}(n_{{\Psi_{\mathcal A}^{\rm imp}}})/\partial n$ should in principle
enable to recover the exact orbitals occupancy in both the bath and the
impurity and, through the functional, the FCI energy. The impurity
Hamiltonian can also be viewed as an embedded version of {Dyall's
Hamiltonian}~\cite{dyallh0}.
 A numerical validation of
Eq.~(\ref{eq:sc_cassoft}) for small molecules can be achieved when
computing from the input FCI orbitals occupation $n$ the
Legendre--Fenchel transform
\begin{eqnarray}
F_{ \mathcal{A}}^{\rm imp}(n)=
\underset{{\varepsilon}
}{\rm sup}\Big\{
\mathcal{E}_{ \mathcal{A}}^{\rm
imp}({\varepsilon})-({\varepsilon}\vert n)
\Big\},
\end{eqnarray}
where $\mathcal{E}_{ \mathcal{A}}^{\rm imp}({\varepsilon})$ is the
ground-state energy of $
\hat{\mathcal{T}}^{\rm HF}+\hat{\mathcal
W}_{\mathcal A}+\sum_p{\varepsilon}_p\hat{n}_p
$. Work is in
progress in this direction.\\

Let us stress that, in contrast to the short-range density functional used in
range-separated DFT, the complementary bath orbital-occupation functional
$\overline{E}^{\rm bath}_{\rm c}(n)$ is not a universal functional. It
will {\it a priori} depend on the molecular orbital basis in which CASSOFT is
formulated. It also depends on the active orbital space. 
Even though range-separated DFT is much simpler in that respect, the
choice of an optimal range-separation parameter is, in terms of
accuracy, not universal partly because static correlation is not a
purely long-range effect. Local and semi-local approximations to the short-range
density functional are usually not accurate enough for modeling strongly
multi-configurational systems (see Ref.~\cite{JChemPhys139_2013} and
the references therein). Working in the orbital space will at least allow for a
proper separation of static and dynamical correlation effects.    
One may
consider the development
of approximate functionals to be a cumbersome task in this context. A
similar challenge is to some extent encountered in {\it Natural Orbital
Functional Theory} (NOFT)~\cite{QUA:QUA24663}. Recent advances in the
field might be useful for such developments. This is left for future
work.\\    

Returning to exact CASSOFT, an AC formula can actually be derived for the complementary bath correlation
functional $\overline{E}^{\rm bath}_{\rm c}(n)$ by introducing
orbital-dependent active spaces ${\mathcal A}_{r,s}$ where the
indices $r$ and $s$ refer to the lowest and highest active orbitals in energy, respectively. Let ${\mathcal A}_{r_0,s_0}$ denote the
active space of interest $\mathcal{A}$. We then consider the auxiliary equations 
\begin{eqnarray}
&&\Bigg(\hat{\mathcal{T}}^{\rm HF}+
\hat{\mathcal
W}_{{\mathcal A}_{r,s}}
+\lambda 
 \hat{\mathcal
W}^{\pm}_{{\mathcal A}_{r,s}}
+\sum_p{\varepsilon}^{\lambda,rs,\pm}_p\,\hat{n}_p
\Bigg)\vert\Psi^{\lambda,rs,\pm}\rangle
=
\mathcal{E}^{\lambda,rs,\pm}\vert\Psi^{\lambda,rs,\pm}\rangle,
\end{eqnarray}
where the orbitals occupation constraint $n_{\Psi^{\lambda,rs,\pm}}=n$
is fulfilled for $0\leq \lambda\leq 1$, $0\leq r\leq r_0$ and $s_0\leq
s$. The
virtual and inactive increment operators are defined as
\begin{eqnarray}
 \hat{\mathcal
W}^{+}_{{\mathcal A}_{r,s}}
=
\hat{\mathcal
W}_{{\mathcal A}_{r,s+1}}
-
\hat{\mathcal
W}_{{\mathcal A}_{r,s}}
,\\
 \hat{\mathcal
W}^{-}_{{\mathcal A}_{r,s}}=
\hat{\mathcal
W}_{{\mathcal A}_{r-1,s}}
-
\hat{\mathcal
W}_{{\mathcal A}_{r,s}},
\end{eqnarray}
respectively. The explicit expression for 
$ 
\hat{\mathcal
W}_{{\mathcal A}_{r,s}}
$
is deduced from Eq.~(\ref{eq:fluctu_pot_active}).  The superscripts
"$+$" and "$-$" refer to the incorporation into the active space ${\mathcal
A}_{r,s}$
of the virtual $s+1$ and inactive $r-1$ orbitals, respectively. A graphical representation is given in
Fig.~\ref{fig:active_space_graph_rep}.
The embedded active electrons are recovered along the AC when $r=r_0$, $s=s_0$ and
$\lambda=0$. According to Eqs.~(\ref{eq:part_F_imp_dyn}) and (\ref{eq:Fimp_LL_qc}), the complementary bath correlation functional can be expressed as
\begin{eqnarray}
\overline{E}^{\rm bath}_{\rm c}(n)&=&F(n)-F^{\rm imp}_{
\mathcal{A}_{r_0,s_0}}(n)\nonumber\\
&=&
\sum_{s\geq s_0}F^{\rm imp}_{\mathcal{A}_{0,s+1}}(n)-F^{\rm
imp}_{\mathcal{A}_{0,s}}(n)
\nonumber\\
&&-
\sum^{r_0}_{r=1}F^{\rm imp}_{\mathcal{A}_{r,s_0}}(n)-F^{\rm
imp}_{\mathcal{A}_{r-1,s_0}}(n),
\end{eqnarray}
thus leading, by analogy with Eq.~(\ref{eq:Hxc_fun_AC_alternative}), to
\begin{eqnarray}\label{eq:AC_bath_cassoft}
\overline{E}^{\rm bath}_{\rm c}(n)&=&
\sum_{s\geq s_0}
\int^1_0 \ddroit \lambda \,\langle \Psi^{\lambda,0s,+}\vert
 \hat{\mathcal
W}^{+}_{{\mathcal A}_{0,s}}\vert \Psi^{\lambda,0s,+}\rangle
\nonumber\\
&&+
\sum^{r_0}_{r=1}
\int^1_0 \ddroit \lambda \,\langle \Psi^{\lambda,rs_0,-}\vert
 \hat{\mathcal
W}^{-}_{{\mathcal A}_{r,s_0}}\vert \Psi^{\lambda,rs_0,-}\rangle.
\end{eqnarray}
It is readily seen from Eq.~(\ref{eq:AC_bath_cassoft}) that the
complementary bath
correlation functional describes correlation effects that (i) couple 
inactive or active orbitals with the virtual orbitals (first term on the
right-hand side) and that
(ii) couple inactive orbitals with active orbitals (second term). The correlation effects
within the active space are treated explicitly. Note that CASSOFT is in
principle exact (in a given basis set) and it is free from 
double counting problems. The reason is that it relies on an exact separation of the LL
functional in the orbital space.
\section{Conclusions and perspectives}\label{sec:conclusions}

The exact formulation of multi-configuration density-functional theory
has been discussed. The infamous double counting problem can be avoided
when separating correlation effects either in the coordinate space or in
the orbital space. In the latter case, orbitals
occupation should be used as basic variable rather than the electron density. This
approach has been applied to the Hubbard Hamiltonian, thus leading to an
exact {\it Site Occupation Embedding Theory} (SOET). The connection with
{\it Density Matrix Embedding Theory} (DMET) has been discussed and an adiabatic
connection (AC) formula has been derived for the complementary bath
functional. The AC has been constructed analytically for the simple
two-site Hubbard model. The computational implementation of SOET as well
as the development of approximate local bath functionals is left
for future work.\\    
We then proposed to apply SOET to a quantum chemical Hamiltonian, thus
showing that multi-configuration methods can be merged
rigorously with orbital-occupation functionals. 
The method is referred to as CAS {\it Site Occupation Functional Theory}
(CASSOFT). In this context,
impurity sites correspond to the active orbitals while inactive and
virtual orbitals are the bath. An AC formalism has also been derived.
The latter should be useful for developing approximate functionals.
A connection with multi-reference perturbation theory may be achieved
from a perturbation expansion of the AC integrand. 
Note also that, in regular DFT, the uniform electron gas played a crucial
role in the development of electron density functionals. 
By analogy, model Hamiltonians such as the Hubbard Hamiltonian could be used
for developing local and semi-local orbital-occupation functionals. Finally, following
Knizia and Chan~\cite{PhysRevLett.109.186404}, it would be interesting to explore connections between CASSOFT and
{\it Density Matrix Renormalization Group} (DMRG)
methods~\cite{PRL92_White_dmrg}. Work is in
progress in these directions.

\section*{Acknowledgments}

The author would like to thank the editorial board for the kind
invitation to submit a "New Views" paper to Mol. Phys.  
EF is grateful to Markus Reiher and Stefan Knecht
for their kind invitation to give a seminar at ETH in august 2014 and
their helpful comments on this work. EF thanks Masahisa Tsuchiizu, Vincent Robert, Laurent Mazouin, Andreas
Savin, Lucia Reining and Bernard Amadon for stimulating discussions on
strongly correlated electrons. The author finally acknowledges financial support from
the PHC program Sakura 2969UK, the LABEX "Chemistry of complex systems"
and the ANR (MCFUNEX project).



\bibliographystyle{tMPH}


\newcommand{\Aa}[0]{Aa}


\label{lastpage}

\end{document}